\documentclass{aa}  

\usepackage{graphicx}
\usepackage{lscape}
\usepackage{txfonts}
\usepackage{enumitem}
\usepackage{multirow}
\newenvironment{sistema}%
{\left\lbrace\begin{array}{@{}l@{}}}%
{\end{array}\right.}
\begin{document}
\def\teff{T_{\mathrm{eff}}}

   \title{Age consistency between exoplanet hosts and field stars}

    \author{A. Bonfanti\inst{1,2} \and S. Ortolani\inst{1,2} \and V. Nascimbeni\inst{2}}
    \institute{Dipartimento di Fisica e Astronomia, Università degli Studi di Padova, Vicolo dell'Osservatorio 3, I-35122 Padova, Italy
    \and
    Osservatorio Astronomico di Padova, INAF, Vicolo dell'Osservatorio 5, I-35122 Padova, Italy}
    \date{}

 
  \abstract
   {Transiting planets around stars are discovered mostly through photometric surveys. Unlike radial velocity surveys, photometric surveys do not tend to target slow rotators, inactive or metal-rich stars. Nevertheless, we suspect that observational biases could also impact transiting-planet hosts.}
   {This paper aims to evaluate how selection effects reflect on the evolutionary stage of both a limited sample of transiting-planet host stars (TPH) and a wider sample of planet-hosting stars detected through radial velocity analysis. Then, thanks to uniform derivation of stellar ages, a homogeneous comparison between exoplanet hosts and field star age distributions is developed.}
   {Stellar parameters have been computed through our custom-developed isochrone placement algorithm, according to Padova evolutionary models. The notable aspects of our algorithm include the treatment of element diffusion, activity checks in terms of $\log{R'_{HK}}$ and $v\sin{i}$, and the evaluation of the stellar evolutionary speed in the Hertzsprung-Russel diagram in order to better constrain age. Working with TPH, the observational stellar mean density $\rho_{\star}$ allows us to compute stellar luminosity even if the distance is not available, by combining $\rho_{\star}$ with the spectroscopic $\log{g}$.}
   {The median value of the TPH ages is $\sim5$ Gyr. Even if this sample is not very large, however the result is very similar to what we found for the sample of spectroscopic hosts, whose modal and median values are [3, 3.5) Gyr and $\sim4.8$ Gyr, respectively. Thus, these stellar samples suffer almost the same selection effects. An analysis of MS stars of the solar neighbourhood belonging to the same spectral types bring to an age distribution similar to the previous ones and centered around solar age value. Therefore, the age of our Sun is consistent with the age distribution of solar neighbourhood stars with spectral types from late F to early K, regardless of whether they harbour planets or not. We considered the possibility that our selected samples are older than the average disc population.}
   {}
   \keywords{stars: fundamental parameters, stars: evolution, (stars:) Herztsprung-Russell and C-M diagrams, stars:planetary systems}

  \maketitle
%

\section{Introduction}\label{sec:intro}
%
Computing ages of field stars is very challenging because the age is not a direct observable. Thanks to models, information about the age comes from the composition and evolutionary state of the core of a star, while we are mostly limited to observing the properties at the surface. Several techniques can be applied.

Using the stellar effective temperature $T_{\mathrm{eff}}$ and luminosity $L$ as input values, age can be computed through interpolation in the grids of isochrones (isochrone placement). Instead, gyrochronology (see e.g. \citealt{barnes10k}) is an empirical technique that allows the determination of stellar ages considering that the rotational speed of stars declines with time because of magnetic braking. Asteroseismology (see \citealt{handler13} for a review) is a very promising technique because the individual oscillation frequencies are directly linked to the inner density profile and the sound propagation speed in the stellar core. These frequencies are recovered through detailed analyses and high-precision photometry, which facilitates the determination of very precise, though model-dependent, ages.
If it is not possible to investigate each oscillation mode, asteroseismic studies simply give global parameters i.e. the large frequency separation $\Delta\nu$ and the frequency of maximum power $\nu_{\mathrm{max}}$, which are linked to the stellar mean density $\rho_{\star}$ and surface gravity $\log{g}$ (see e.g. \citealt{kjeldsen95}). In this case, asteroseismology loses part of its strength. Input $T_{\mathrm{eff}}$, $\rho_{\star}$ and $\log{g}$ again require isochrones to compute ages as in \citet{chaplin14}, however, asteroseismic $\log{g}$ is known with better precision if compared with the spectroscopic value, for instance. For a broad review about different age computation methods, see \citet{soderblom10}.

Since $T_{\mathrm{eff}}$ and $L$ can be usually recovered for many stars, in this paper we compute the ages of transiting-planet host stars (TPH) in a homogeneous way via isochrones. Knowledge of stellar ages is particularly important in the context of planet-hosting stars (SWP). The age distribution of SWP tells us whether planets are preferentially hosted by young or old stars. This is related to the dynamical stability of the systems and with the mutual influence between planets and hosting star; see e.g. \citet{paetzold04}, \citet{barker09}, \citet{debes10}. Moreover, ages for exoplanet host stars enable a comparison with typical timescales of biological evolution and an assessment of the plausibility of the presence of life (see e.g. \citealt{kasting03}).

The paper is organised in the following way: §\ref{sec:data} describes the characteristics of our stellar sample and the isochrones. §\ref{sec:method} presents the central aspects of our algorithm, §\ref{sec:results} shows the results, and §\ref{sec:conclusion} summarizes our work.
   

\section{The data}\label{sec:data}
\subsection{Planet-hosting stars catalogues}
We analysed the ages of those stars whose planets were discovered through the transit method. In principle, these kinds of stars should not suffer from biases: (1) The stars that are chosen are not necessarily inactive, unlike in radial velocity surveys, where spectroscopic analysis requires sharp and well-defined lines. However, we caution that it is indeed more difficult to detect transits for stars with a large amplitude of intrinsic variability. (2) These TPH stars are not necessarily slow rotators, unlike in radial velocity surveys. In fact, rotation broadens the lines and reduces their depth, making spectroscopic analysis less precise, however, once a possible transit signal is detected, spectroscopic validation is required to confirm such a planet. Therefore, stars belonging to photometric surveys must also be suitable for spectroscopic analyses if exoplanet validation is expected, so almost the same biases are expected.
In fact, in the case of transiting planet hosts, there are other systematic selection effects.
Transiting-planet hosts are expected to be preferentially edge-on, even if spin-orbit misalignment occurs in some exoplanetary systems. Gravity darkening or differential rotation (\citealt{vonZeipel24}, \citealt{maeder99}) could affect stellar observables. In addition, the hosted planets are very close to their own star.

We selected 61 transiting-planet hosts from SWEET-Cat, a catalogue of stellar parameters for stars with planets\footnote{%
\url{https://www.astro.up.pt/resources/sweet-cat/}} \citep{santos13}%
, to obtain our transiting-planet hosts (TPH) catalogue. Among the stars of this catalogue, we further consider only those stars brighter than $V=12$ and this inevitably introduces a further source of bias. This criterion takes into account that future photometric missions with the aim of characterizing exoplanets, such as CHEOPS \citep{broeg13} or PLATO \citep{rauer14}, will investigate bright stars. This led us to the Bright Transiting-Planet Hosts (BTPH) catalogue, which is composed of 43 stars.
The metallicity [Fe/H] and the logarithm of the surface gravity $\log{g}$ are always available from Sweet-Cat. If available, we took $V$ magnitude and $B-V$ colour index from \citet{maxted11}, otherwise we collected $V$ from SWEET-Cat and $B-V$ from The Site of California and Carnegie Program for Extrasolar Planet Search: Exoplanets Data Explorer.\footnote{%
\url{http://exoplanets.org/table}} %
As reported by \citet{maxted11}, the target stars of surveys that aim to discover exoplanets through transits are typically characterized by optical photometry of poor quality in the range V=8.5-13 mag. For stars brighter than $V\approx12$, optical photometry is usually available from Tycho catalogue, nevertheless, this catalogue is only complete up to $V\approx11$ and photometric accuracy rapidly deteriorates for $V\gtrsim9.5$. The authors give high-quality photoelectric optical photometry for planet-hosting stars (mostly WASP discoveries), so we decided to use these data if available.

We also built a catalogue of 274 planet-hosting stars whose planets were detected through radial velocity method (Spectroscopic hosts: SH catalogue) from SWEET-Cat.

\subsection{Solar neighbourhood catalogues}\label{subsec:SNcat}
We built a catalogue of F-G-K main sequence stars (MS-stars) belonging to the solar neighbourhood (SN catalogue) by taking stellar data from the re-analysis of the Geneva-Copenhagen survey by \citet{casagrande11}. It is a survey of late-type dwarf stars that are magnitude limited at $V\approx8.3$; the authors computed the ages for these stars. In particular, we collected the 7044 stars with available ages, belonging to the MS. The MS containing the F-G-K stars has been indentified by selecting a strip going from $T_{\mathrm{eff}}\approx4500$ K to $T_{\mathrm{eff}}\approx7100$ K, within a range of 0.45 dex in $\log{L}$, whose minimum and maximum values are $-1.24$ dex and $0.63$ dex, respectively. We further removed F type stars, i.e stars with $\teff>6300$ K, from the SN catalogue. This way we remained with 3713 stars (Reduced Solar Neighbourhood catalogue; RSN) belonging to the same spectral type range of planet-hosting stars. 
Among useful input parameters to compute stellar ages through our own algorithm, \citet{casagrande11} give only metallicity, which is inferred from Str\"{o}mgren photometry; distance, according to the new reduction of the \emph{Hipparcos} parallaxes \citep{vanLeeuwen07}; and V magnitude for each star. We complemented this information by cross-matching the entire Geneva-Copenhagen survey with the catalogue of cool late-type stars by \citet{valenti05}, which also provides precise spectroscopic measurements of surface gravity $\log{g}$ and projected rotational velocity $v\sin{i}$. This led to the Valenti Fischer Solar Neighbourhood catalogue (VF-SN catalogue), which contains 825 stars.

\begin{figure}
 \centering
 \includegraphics[width=\columnwidth]{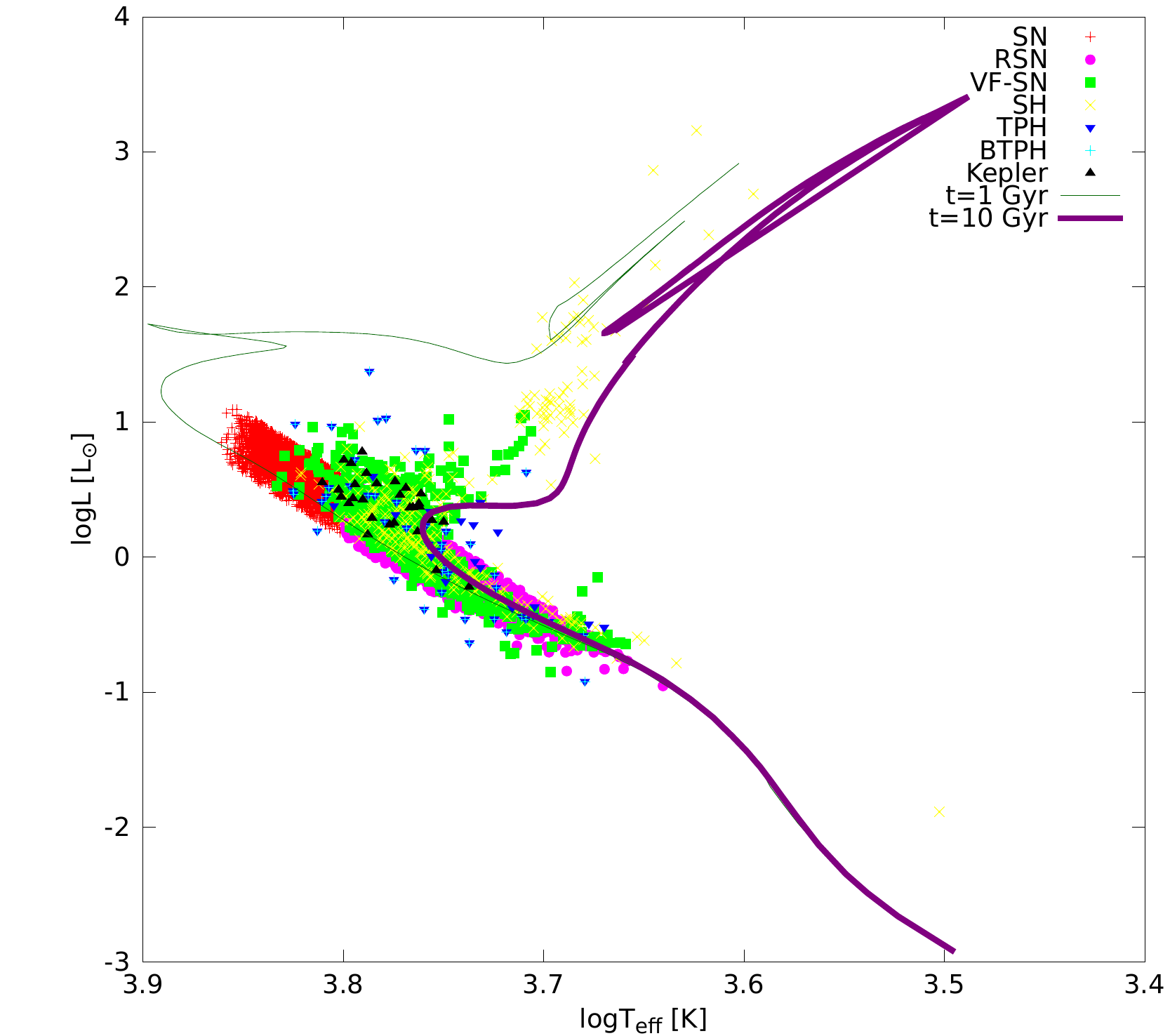}
 \caption{Stars belonging to our custom-built catalogues are represented on the HRD. Two solar metallicity isochrones, corresponding to 1 Gyr and 10 Gyr, are shown as reference. Since BTPH is a subsample extracted from TPH, all the cyan crosses representng the BTPH stars are superimposed on part of the blue reverse triangles representing the TPH stars.}
 \label{fig:HRDsamples}
\end{figure}

A brief overview of our custom-built catalogues used in the paper is given in Table \ref{tab:catalogs}. In Fig. \ref{fig:HRDsamples} stars, belonging to our catalogues are represented on the HRD with two solar metallicity isochrones as reference.

\begin{table*}
\caption{Overview of our custom-built catalogues}             
\label{tab:catalogs}      
\centering          
\begin{tabular}{lcll}
\hline\hline       
                     
Catalogue & \# stars & $\log{g}$ source & Reference \\ 
\hline                    
TPH: Transiting planet hosts 	& 61 	& spectroscopy 	& SWEET-Cat \\
BTPH: Bright transiting planet hosts 	& 43	& spectroscopy 	& SWEET-Cat \\
SH: Spectroscopic hosts 	& 274	& spectroscopy 	& SWEET-Cat \\
SN: Solar neighbourhood 	& 7044	& not available	& \citet{casagrande11} \\
RSN: Reduced solar neighbourhood	& 3713	& not available	& \citet{casagrande11} \\
\multirow{2}*{VF-SN: Valenti Fischer solar neighbourhood}& \multirow{2}*{825}	& \multirow{2}*{spectroscopy}	& \citet{casagrande11} + \\
                               &       &              & \citet{valenti05} \\
Kepler sample\tablefootmark{a} & 29 & asteroseismology & \citet{silvaAguirre15} \\
\hline                  
\end{tabular}
\tablefoottext{a}{see §\ref{subsec:testAlgorithm}}
\end{table*}

\subsection{Isochrones}
To compute the ages of stars we used isochrones taken from Padova and Trieste Stellar Evolutionary Code (PARSEC, version 1.0)\footnote{\url{http://stev.oapd.inaf.it/cgi-bin/cmd}} by \citet{bressan12}. We queried isochrones identified by $\log{t}$ in the range between 6 and 10.1 ($t$ in years) at steps of 0.05 dex. These isochrones include the pre-MS phase, so the given ages must be considered as starting from the birth of a star and not since the zero age main sequence (ZAMS). Specific details about the solar parameters adopted by the isochrones are already reported by \citet{bonfanti15}. Here, we recall the relation between metallicity $Z$ and [Fe/H], i.e.
\begin{equation}
 Z=10^{\mathrm{[Fe/H]}-1.817}
 \label{eq:Z_FeH}
\end{equation}

\section{Age determination methods}\label{sec:method}
Computing the age of a field star through isochrones requires us to put the star on a suitable plane with its error bars. Traditionally, HRD is chosen, so $T_{\mathrm{eff}}$ and $L$ are the reference quantities. Several catalogues in the literature already report $\teff$ or $L$, but they were obtained by different authors through different processes and/or calibration techniques. For instance, $\teff$ and $L$ are not likely to be consistent with the colour-temperature scale or the bolometric corrections (BCs) adopted by the isochrones. Therefore, we prefer to start from quantities coming from observations in a straightforward way, where possible. Our reference input quantities to compute stellar ages are $V$ magnitude, $B-V$ colour index, [Fe/H] metallicity, spectroscopic $\log{g}$, and parallactic distance $d$, which can be substituted by the $\frac{a}{R_{\star}^3}$ parameter coming from transit, as better specified in \S~\ref{subsec:IsocPreliminary}.

\subsection{Isochrone placement: Preliminary considerations}\label{subsec:IsocPreliminary}
Starting from observational quantities, $T_{\mathrm{eff}}$ can be inferred from colour index (e.g. $B-V$), while $L$ is determined thanks to the magnitude in a given band (say $V$), its corresponding bolometric correction $BC_V$ and the distance $d$ of the star recovered from parallax $\pi$. In the particular case where a star hosts a transiting planet, we are able to compute $L$, even if $d$ is not available. In fact, the ratio between the orbital period $P$ and the transit duration allows us to recover $a/R_{\star}$, where $a$ is the planet semimajor axis and $R_{\star}$ is the stellar radius (see e.g. \citet{winn10}). Rearranging Kepler III law in the manner shown by \citet{sozzetti07}, the mean stellar density results to be
\begin{equation}
 \rho_{\star}=\frac{3\pi}{G}\left(\frac{a}{R_{\star}}\right)^3\frac{1}{P^2}
 \label{eq:rhostar}
\end{equation}
where $G$ is the universal gravitational constant.
Combining the spectroscopic $\log{g}$ with $\rho_{\star}$, one can solve a system of two equations in the two variables $R_{\star}$ and $M_{\star}$. One obtains
\begin{equation}
 \begin{sistema}
  \frac{R_{\star}}{R_{\odot}}=\frac{g}{g_{\odot}}\left(\frac{\rho_{\star}}{\rho_{\odot}}\right)^{-1} \\
  \frac{M_{\star}}{M_{\odot}}=\left(\frac{g}{g_{\odot}}\right)^3\left(\frac{\rho_{\star}}{\rho_{\odot}}\right)^{-2}
 \end{sistema}
 \label{eq:R&M}
\end{equation}
Finally, the stellar luminosity $L$ is given by
\begin{equation}
 \frac{L}{L_{\odot}}=\left(\frac{R}{R_{\odot}}\right)^2\left(\frac{\teff}{T_{\mathrm{eff},\odot}}\right)^4
 \label{eq:L}
\end{equation}

\begin{figure}
 \centering
 \includegraphics[width=\columnwidth]{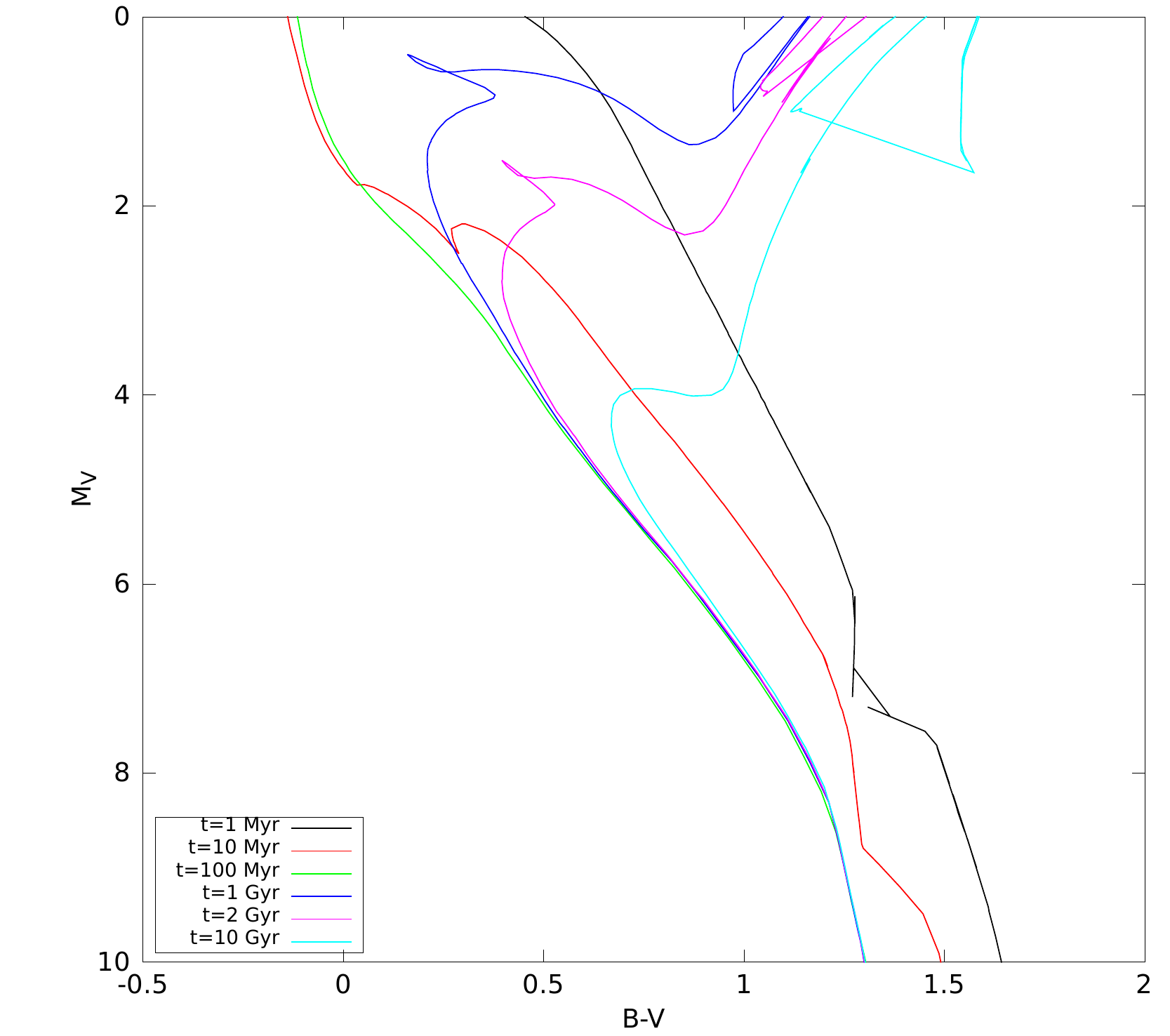}
 \caption{Isochrones of solar metallicity: pre-MS isochrones are located on the right-hand side of the MS (in this region ages increase from right to left on the diagram) and they intersect older isochrones around the TO. The ages of MS isochrones increase from left to right on the diagram.}
 \label{fig:TOintersection}
\end{figure}

In \citet{bonfanti15} we have already pointed out that on the right-hand side of the main sequence (in the lower temperature region) or around the turn-off (TO) very old and very young isochrones are close and can even overlap. According to Fig. \ref{fig:TOintersection}, the 1-Myr- and 10-Myr-isochrones interesect older isochrones in the TO region. The intersection points are representative of a degeneracy between pre-MS and MS isochrones on the CMD. In fact, the photometry alone cannot disentagle young and old ages, and other information is needed.

So far, the only exoplanet candidate orbiting around a pre-MS is PTFO 8-8695b, as reported by \citet{vanEiken12} and then investigated by \citet{barnes13}. Thus, we do not expect to find pre-MS stars among our samples of stars with planets. Anyway, our algorithm is built to compute ages of any kind of star and we decided to perform the activity checks that are described in the following. In this way, we do not put any a priori conditions on the evolutionary stage of the planet-hosting stars. Possible pre-MS interlopers in our planet-hosting stars samples would be very low-mass stars with long pre-MS lifetimes. Twenty-two stars out of the 335 SWP have masses lower than 0.8 $M_{\odot}$. In principle, they could be pre-MS stars and the checks performed by our algorithm may help in recognizing them.

\begin{figure}
 \centering
 \includegraphics[width=\columnwidth]{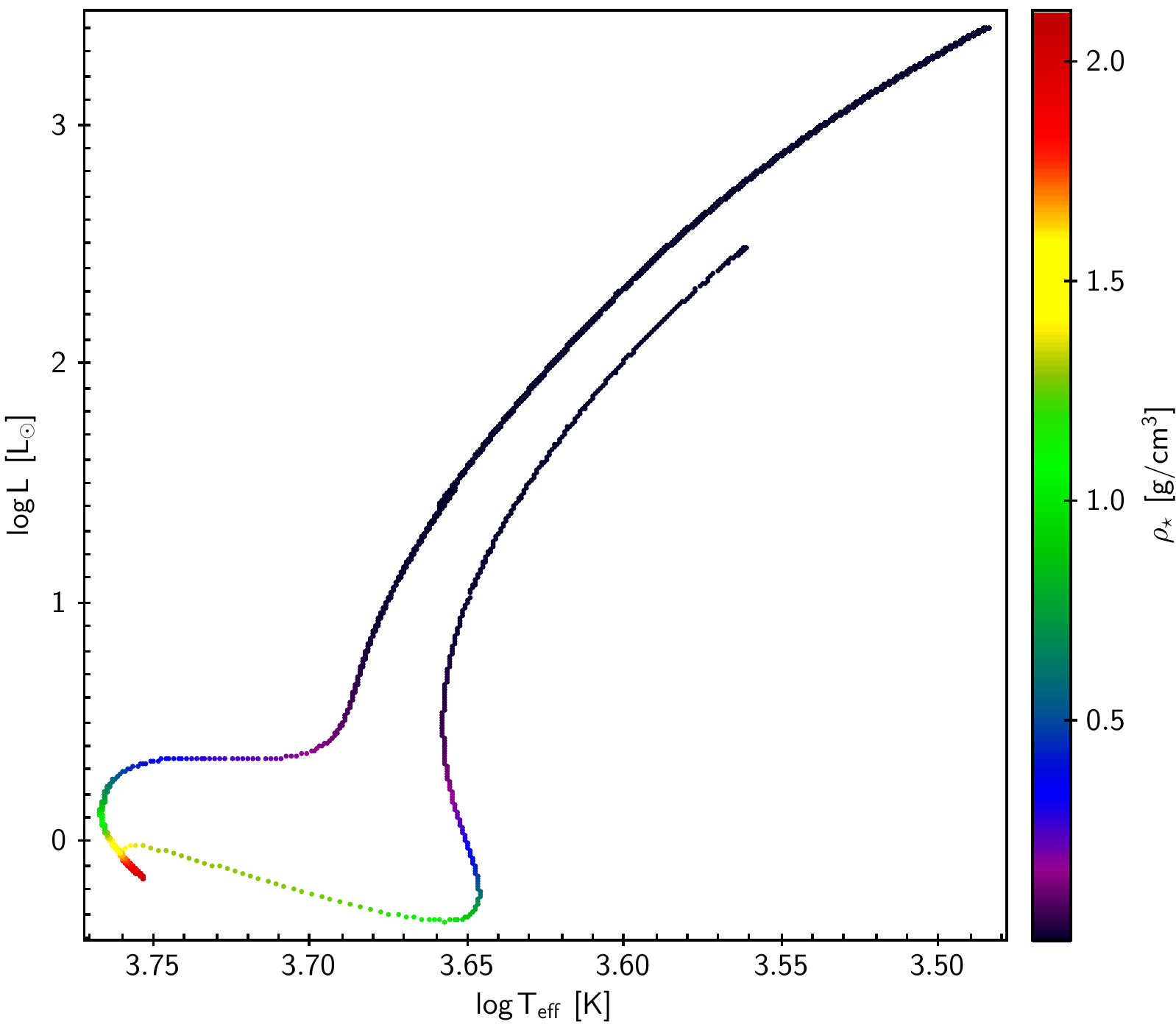}
 \caption{Evolutionary track of a star with $M=1M_{\odot}$ and $Z=Z_{\odot}$. Even if the oldest MS isochrones are close to pre-MS isochrones, however, $\rho_{\star}$ in the MS is sensibly higher with respect to the pre-MS phase. As a consequence, the mean stellar density enables us to discard unlikely age values according to $\rho_{\star}$.}
 \label{fig:track_rhoEvo}
\end{figure}

In the case of late spectral type stars, such as those analysed in this paper, very young stars are chromospherically very active with respect to older stars and typically rotate faster, so we performed activity checks in terms of $\log{R'_{HK}}$ and $v\sin{i}$, trying to remove the degeneracy between pre-MS and MS isochrones. We evaluated three age scales through three independent methods to decide on the ensemble of isochrones to be used in the following computation.
\begin{enumerate}
\item We considered the age-activity relation by \citet{mamajek08} and we set a conservative threshold of $~0.2$ dex corresponding to the typical difference between the highest and lowest peaks in activity and the average level for a solar-type star. Inserting $\log{R'_{HK}}$ in the relation, we evaluated the corresponding age: $\tau_{HK}$ represents this age if it was younger than 500 Myr, otherwise $\tau_{HK}=500$ Myr.
\item\label{it:vsini} \citet{meibom15} proved that the gyrochronological relation by \citet{barnes10} holds up to 2.5 Gyr, so we applied this relation employing $\frac{4}{\pi}v\sin{i}$ as the expected stellar rotational velocity to obtain the gyro age $\tau_{v}$. We set $\tau_v=2.5$ Gyr, if the resulting gyro age was older than 2.5 Gyr. 

\item There is other information that may suggest whether a star located under the TO on the right-hand side of the MS is very young or very old, and this is $\rho_{\star}$. Figure \ref{fig:track_rhoEvo} shows the evolutionary track of a 1 $M_{\odot}$ star with solar metallicity. Starting from the birth of a star, $\rho_{\star}$ increases in approaching the MS. After the TO, $\rho_{\star}$ clearly decreases so that post-MS stars have a mean stellar density similar to that of pre-MS stars. So, for $M_V>5$, corresponding to the luminosity of the TO of the oldest isochrone in the CMD, pre-MS isochrones differ from older isochrones in terms of $\rho_{\star}$. Among pre-MS ages, $\tau_\rho$ is the threshold age value such that for $t<\tau_\rho$ isochrones report mean stellar density $\rho<\rho_{\star}$.
\end{enumerate}

The maximum value among $\tau_{HK}$, $\tau_v$ and $\tau_{\rho}$ represents the age up to which all the younger isochrones are discarded before the computation of stellar age. 

\subsection{Isochrone placement: Implementation}
The isochrone placement technique enables the determination of the ages of field stars, as well as all the other stellar parameters, such as $T_{\mathrm{eff}}$, $L$, $\log{g}$, $M_{\star}$, $R_{\star}$, according to stellar evolutionary models.
This technique was already described in \citet{bonfanti15}, but several improvements have been made, such as the new kind of activity checks described above and the possibility of computing the age without any input distance $d$, if we have stellar density measurements. We also solved some problems linked to numerical stability convergence for which the previous algorithm sometimes gave fictitious young ages. In fact, in the previous version of the algorithm some input data were loaded in single precision, instead of double precision. Sometimes, it could happen that single precision were not sufficient to perform the correct computation of stellar parameters. Moreover, we make this new version more flexible, since it also enables to use input asteroseismic global parameters or only spectroscopic parameters if photometry is not available. Here, we briefly summarize the key aspects.
 
To make as few assumptions as possible and to start from input data directly obtained from observations, our algorithm requires
\begin{itemize}
 \item visual magnitude $V$;
 \item colour index $B-V$;
 \item metallicity [Fe/H];
 \item spectroscopic $\log{g}$;
\end{itemize}

and the distance $d$ or $a/R_{\star}$. If $d$ is available, it is possible to infer $T_{\mathrm{eff}}$ from $B-V$ and $L$ from the absolute magnitude $M_V$ via interpolation in the isochrone grid. Then $R_{\star}$ is known thanks to Stefan-Boltzmann law (Eq. 4) and finally $M_{\star}$ can be computed by combining $R_{\star}$ with $\log{g}$. If, instead, $d$ is not available, which can occur for some TPH, first we compute $\rho_{\star}$ through (\ref{eq:rhostar}) and then we recover $M_{\star}$ and $R_{\star}$ via relations (\ref{eq:R&M}). After that, we obtain the correspondence between $B-V$ and $T_{\mathrm{eff}}$ through interpolation in the isochrone grid. Finally, we compute $L$ from $T_{\mathrm{eff}}$ with (\ref{eq:L}).
Once all the input parameters are available, it is possible to derive stellar properties according to Padova theoretical models by properly weighting each isochrone in the manner already described in \citet{bonfanti15}. With the new version of the algorithm, we improved the weighting scheme to take the evolutionary speed of a star into account. In fact, the probability that a star is a given age does not only depend on the simple distance between the star and the given isochrone in the HRD, but it is also influenced by the time spent by a star in a given evolutionary stage. For instance, pre-MS evolution is quicker than the MS one. This means that a star rapidly changes its properties during the first tenths of Myr of its life, instead, it remains in the MS for Gyrs with parameters variations detectable on longer timescales.
As a consequence, given a star on the HRD, located at the same distance with respect to a pre-MS and a MS isochrone, the probability of dealing with a MS star is higher. To quantify this aspect, we considered the theoretical stellar evolutionary track characterized by the same input metallicity and mass of the star, and we evaluated its evolutionary speed by
\begin{equation}
 v_{\mathrm{evo}}=\sqrt{\left(\frac{\log{L_2}-\log{L_1}}{t_2-t_1}\right)^2+\left(\frac{\log{T_{\mathrm{eff,2}}}-\log{T_{\mathrm{eff,1}}}}{t_2-t_1}\right)^2}
 \label{eq:vevo}
\end{equation}
where ($\log{T_{\mathrm{eff,1}}}$, $\log{L_1}$) is the point on the track, that is nearer to the star, while ($\log{T_{\mathrm{eff,2}}}$, $\log{L_2}$) is the point that occurs later in time on the track, and $t_1$ and $t_2$ are the epochs reported by the track. The greater $v_{\mathrm{evo}}$, the less is the probability to find a star in such an evolutionary stage. We normalized $v_{\mathrm{evo}}$, with respect to a reference speed value $v_{\mathrm{ref}}$ for a given track, that is the lowest speed registered on the entire track. In this way, the evolutionary speed can be easily interpreted as a multiple of a reference speed with which a star goes along its track and the contribution to be added to the weight is unitless, like the others. So the weight $p_i$ to be attributed to the $i^{\mathrm{th}}$ isochrone results to be

\begin{multline}
p_{i}=\left[\left(\frac{\log{L}-\log{L_i}}{\Delta\log{L}}\right)^2+\left(\frac{\log{T_\mathrm{eff}}-\log{T_{\mathrm{eff,}i}}}{\Delta\log{T_{\mathrm{eff}}}}\right)^2+\left(\frac{M_{\star}-M_{\star,i}}{\Delta M_{\star}}\right)^2+\right. \\
+\left. \left(\frac{\log{g}-\log{g_i}}{\Delta\log{g}}\right)^2+\log^2{\left(\frac{v_{\mathrm{ref}}}{v_{\mathrm{evo}}}\right)}\right]^{-1}
\label{eq:peso}
\end{multline}

Sometimes photometry is not available and only spectroscopic analyses have been carried out. We caution that the given spectroscopic input temperature has been inevitably subjected to a calibration process, which can bring biases. Anyway, to compute ages in such cases we can use spectroscopic [Fe/H], $T_{\mathrm{eff}}$ and $\log{g}$. In this case, the algorithm works in the $\log{g}$-$\log{T_{\mathrm{eff}}}$ plane following the same prescriptions as in the HRD. This time the weight is simply given by
\begin{equation}
p_{i}=\left[\left(\frac{\log{g}-\log{g_i}}{\Delta\log{g}}\right)^2+\left(\frac{\log{T_\mathrm{eff}}-\log{T_{\mathrm{eff,}i}}}{\Delta\log{T_{\mathrm{eff}}}}\right)^2+\log^2{\left(\frac{v_{\mathrm{ref}}}{v_{\mathrm{evo}}}\right)}\right]^{-1}
 \label{eq:peso_rid}
\end{equation}
where this time the evolutionary speed of the star $v_{\mathrm{evo}}$ and its reference value $v_{\mathrm{ref}}$ are evaluated in the $\log{g}$-$\log{\teff}$ plane instead of the HRD.

If global asteroseismological indexes, i.e. $\Delta\nu$ and $\nu_{\mathrm{max}}$, are available, $\log{g}$ and $\rho_{\star}$ may also be computed by inverting the following scaling relations:
\begin{equation}
 \Delta\nu=\sqrt{\frac{M_{\star}}{M_{\odot}}\left(\frac{R_{\star}}{R_{\odot}}\right)^{-3}}\Delta\nu_{\odot}
 \label{eq:DeltaNu}
\end{equation}
\begin{equation}
 \nu_{\mathrm{max}}=\frac{g}{g_{\odot}}\left(\frac{\teff}{T_{\mathrm{eff},\odot}}\right)^{-\frac{1}{2}}\nu_{\mathrm{max},\odot}
 \label{eq:NuMax}
\end{equation}
$\Delta\nu_{\odot}=135.1$~$\mu$Hz and $\nu_{\mathrm{max},\odot}=3090$~$\mu$Hz as reported by \citet{chaplin14}. Knowledge of both $\log{g}$ and $\rho_{\star}$ enables us to compute $M_{\star}$ and $R_{\star}$. Given that $T_{\mathrm{eff}}$ is available, $L$ may also be recovered using (\ref{eq:L}). Even if an accurate asteroseismological analysis based on the study of individual frequency enables precise determination of the stellar evolutionary stage, however, combining information from global asteroseismic parameters and from spectroscopy gives a complete set of input data useful for our isochrone placement. Once the star is located on the HRD, it is then possible to compute its age and its parameters according to Padova evolutionary models. Our algorithm takes element diffusion into account.

If known, stellar multiplicity has been pointed out through a flag at the column \emph{Bin} of Table \ref{tab:exostars}. In these cases, the literature already reports data referred to the specific star we analysed. We caution that if some unresolved binaries were present in our samples, such stars would appear more luminous than they are. If located in the MS region, they would erroneously be judged as older.

Another critical point deals with reddening. In the case of SWP, we do not deeply check whether the different sources give photometry de-reddened or not because neither \emph{SWEET-Cat} nor the \emph{Exoplanets Data Explorer} report any reddening information. We explicitely account for reddening in the case of those TPH listed in \citet{maxted11}, who report the colour excess $E(B-V)$.  Anyway, by a posteriori catalogue cross-matching, we were able to recover $E(B-V)$ index for 154 stars out of the 335 SWP, and we found that more than $80\%$ of them has $E(B-V)=0$. Similarly, $\sim90\%$ of the stars belonging to the VF-SN catalogue have $E(B-V)=0$. Considering also that the analysed stars are essentially inside the Local Bubble, whose extension varies between $\sim80$ and 200 pc from the Sun \citep{sfeir99}, we expect that the effect of reddening does not significantly impact our resulting statistics.

\section{Results}\label{sec:results}
\subsection{Test of the algorithm}\label{subsec:testAlgorithm}
\citet{silvaAguirre15} analysed a sample of Kepler exoplanet host stars (Kepler sample from here on). They performed a complete asteroseismological analysis of the individual oscillation frequencies, recovered thanks to the high signal-to-noise ratio of their observations. Among other fundamental properties, they derived the ages of their sample of stars, claiming a median error of 14\%. As a result of high reliability attributed to a complete asteroseismological analysis, comparing the results given by our isochrone placement with those reported by \citet{silvaAguirre15} represents a good validation test for our algorithm. The authors observe that the majority of these stars are older than the Sun because of selection effects. In particular, stellar pulsations characterized by high signal-to-noise ratio are preferentially detected in F-type stars (ages $\sim2$-3 Gyr) and in old G-type stars (ages $\sim6$ Gyr). Thus, aim of this section is to test the accuracy of our algorithm, without comparing the 
evolutionary stage of the Kepler sample with other stars.

We analysed 29 over 33 stars of the Kepler sample, for which both $\Delta\nu$ and $\nu_{\mathrm{max}}$ were available. We have just considered the global asteroseismic parameters, deriving input $\log{g}$ and $\rho_{\star}$ by inverting (\ref{eq:DeltaNu}) and (\ref{eq:NuMax}). Spectroscopic [Fe/H] and $T_{\mathrm{eff}}$ were reported by \citet{silvaAguirre15}. If available, $v\sin{i}$ was employed to perform checks as described in point \ref{it:vsini} in \S~\ref{subsec:IsocPreliminary}.

\begin{figure}
 \centering
 \includegraphics[width=\columnwidth]{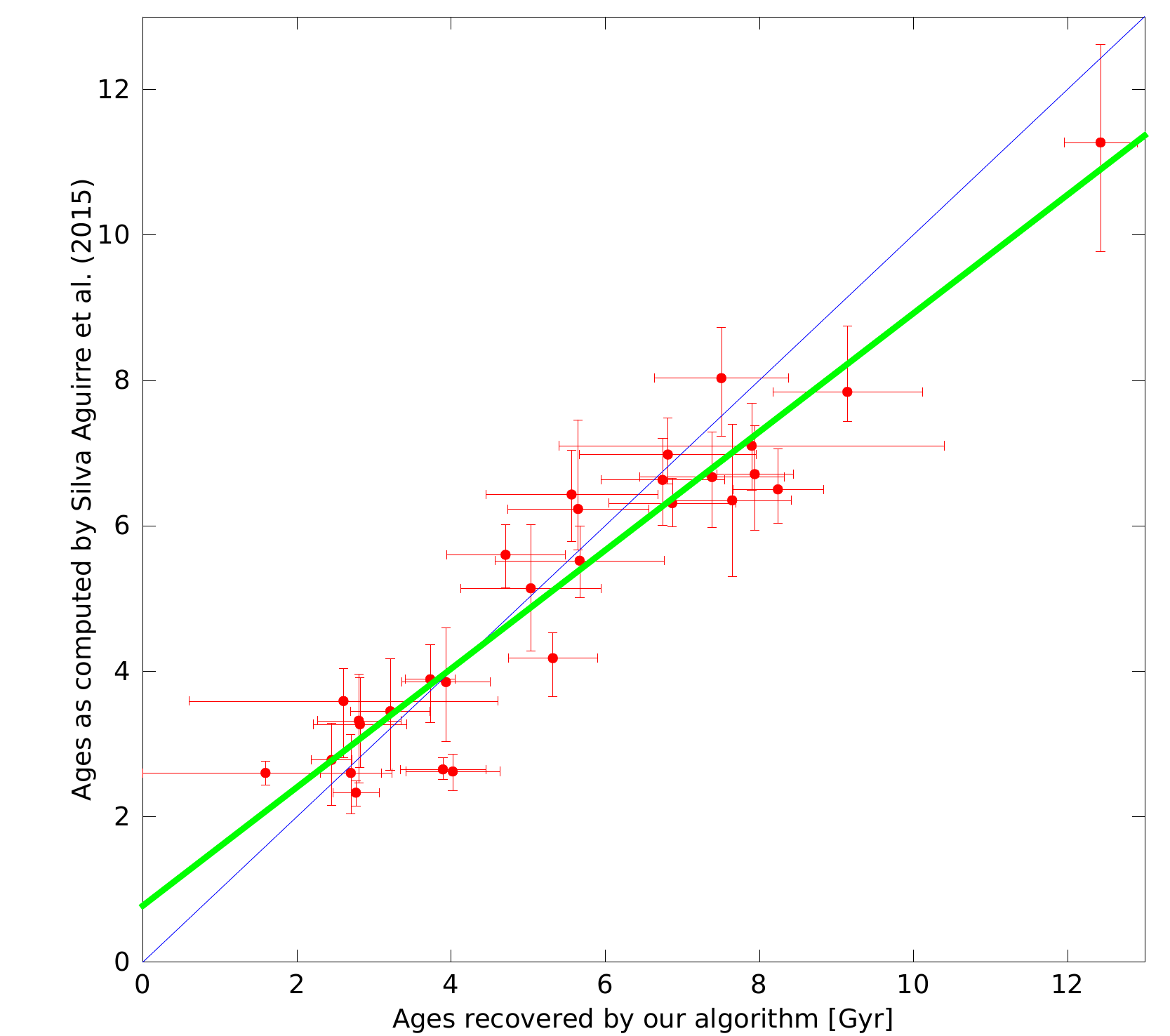}
 \caption{Ages of the Kepler sample. The least-squares line that regress the data is represented with a thick green line, while the thin blue line is the bisector representing the exact correspondence between the data.}
 \label{fig:AgesKepler}
\end{figure}

Our age determination is in good agreement with the analysis of \citet{silvaAguirre15}, as shown in Fig. \ref{fig:AgesKepler}. The linear correlation coefficient $r=0.95$ and the reduced $\chi^2/26=1.5$ confirm that a linear least-squares regression well describe the data scatter and is consistent with the extension of our error bars. The least-squares line represented in green (thicker line) shows that our method slightly overestimate the age in the domain of the oldest stars.

\subsection{The ages of the exoplanet hosts}
\begin{figure}
 \centering
 \includegraphics[width=\columnwidth]{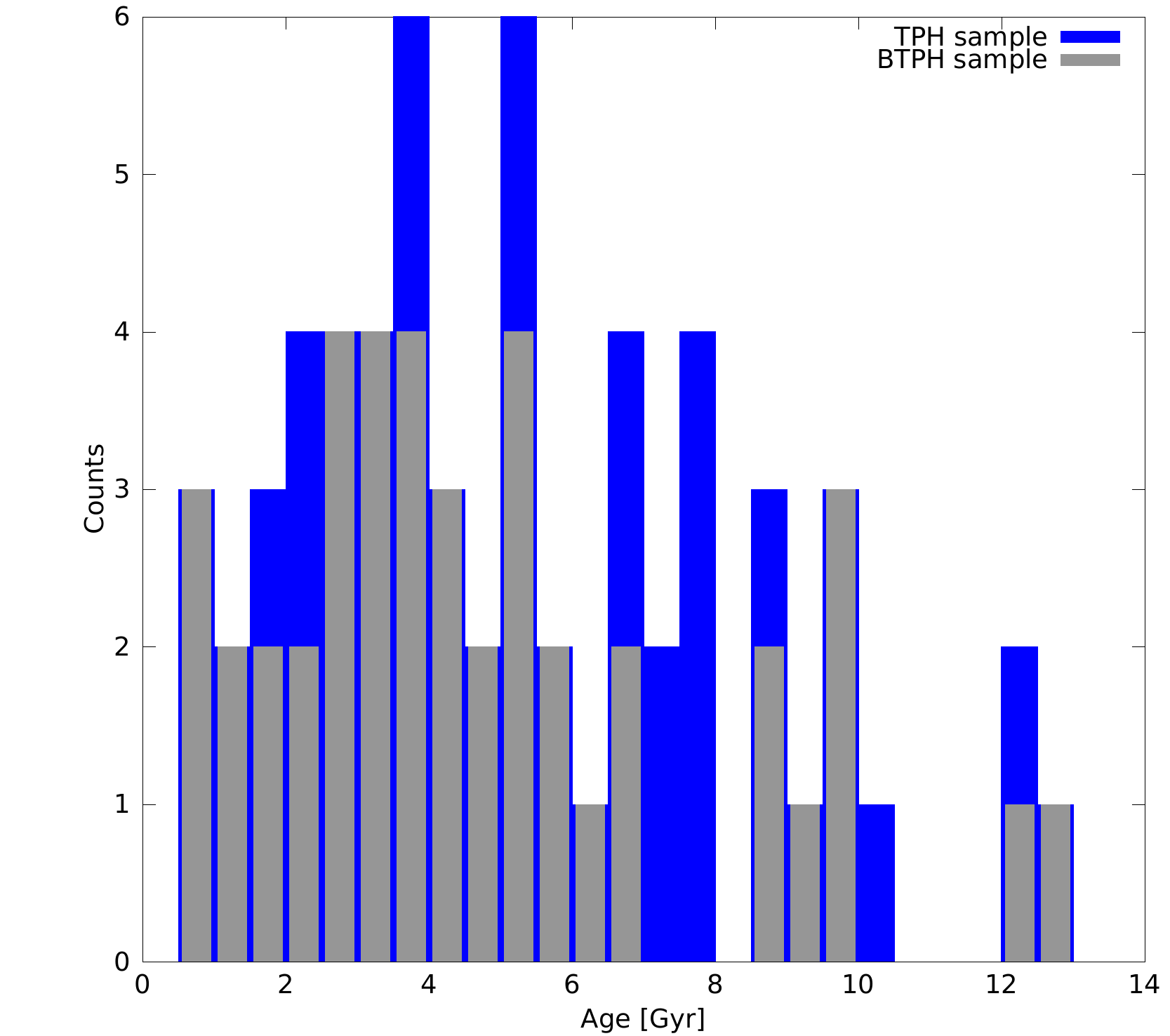}
 \caption{Grey bars represent age distribution of the 43 stars belonging to the BTPH catalogue. The superset given by TPH catalogue is represented in the background through blue bars.}
 \label{fig:AgeTPH}
\end{figure}

The BTPH catalogue is a subset of the TPH catalogue. In Fig. \ref{fig:AgeTPH}, we superimposed the age distribution of the 43 stars belonging to the BTPH catalogue (grey bars) to the age distribution of the all TPH (blue bars).
The medians of the distributions are $\sim4.2$ Gyr and $\sim5$ Gyr for the BTPH and for the TPH, respectively.
One of the three stars younger than 1 Gyr, namely WASP-18 ($t=0.9\pm0.2$ Gyr), appears too blue for its metallicity, so we investigated the input parameters of this star. \citet{southworth09} analysed the properties of WASP-18, adopting $V=9.30$ and $B-V=0.44$, instead of $V=9.273$ and $B-V=0.484$, which are the values we used. In addition they started from [Fe/H]=0, as reported by \citet{hellier09}, which sensibly differs from [Fe/H]=0.19, which we took from SWEET-Cat. With the input parameters used by \citet{southworth09}, WASP-18 is again located on the bluer side out of the MS, and our isochrone placement gives $t=0.2_{-0.2}^{+0.3}$ Gyr. \citet{southworth09} considered different sets of evolutionary models and they concluded that WASP-18 is age between 0 and 2 Gyr. This is consistent with both of our determinations, but since the different sets of input parameters do not fully agree with Padova theoretical models, we caution that the age estimation is not necessarily reliable. 
Further photometric investigations or a re-determination of its metallicity are required to correct the inconsistency between the position of the star on the HRD and the theoretical isochrones.

\begin{figure}
 \centering
 \includegraphics[width=\columnwidth]{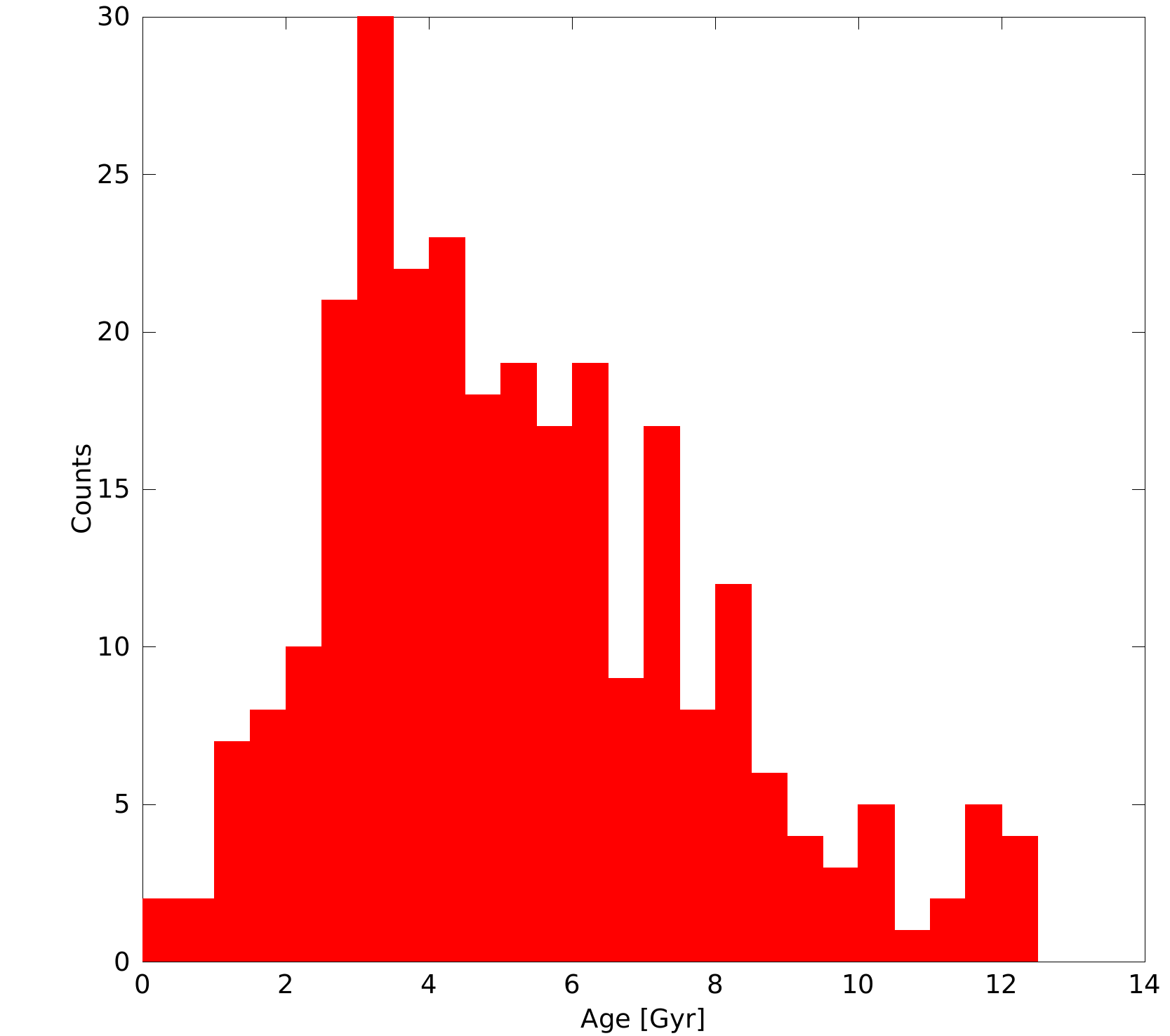}
 \caption{Age distribution of the 274 stars belonging to the SH catalogue. The median of the distribution is $\sim4.8$ Gyr, which is very close to the age of the Sun.}
 \label{fig:AgeSH}
\end{figure}

As a term of comparison, we computed the ages of stars taken from SH catalogue. The consequent age distribution is represented in Fig. \ref{fig:AgeSH}. This age distribution peaks in the [3, 3.5) Gyr bin and its median is $\sim4.8$ Gyr, which is very close to the solar age value. The age distributions in Figs. \ref{fig:AgeTPH} and \ref{fig:AgeSH} are consistent. Differences may arise because of the paucity of the TPH, but, in any case, no significant bias emerges in the comparison between the samples. Actually, we performed a Kolmogorov-Smirnov test (KS test) to investigate whether TPH and BTPH come from the same distribution, which characterizes the larger SH sample.
The high p-values (0.5 for the TPH-SH comparison and 0.3 for the BTPH-SH comparison) suggest that we should not reject the null hypothesis based on which the samples come from the same distribution. This emphasizes that even if photometric and spectroscopic targets could be chosen according to different criteria, the confirmation of a candidate exoplanet requires the application of both the transit and radial velocity method. Therefore, similar biases are expected in the two different samples.

All the parameters of the planet-hosting stars derived according to Padova isochrones are listed in Table \ref{tab:exostars}.


\subsection{Age comparison with the stars of the solar neighbourhood}
\begin{figure}
 \centering
 \includegraphics[width=\columnwidth]{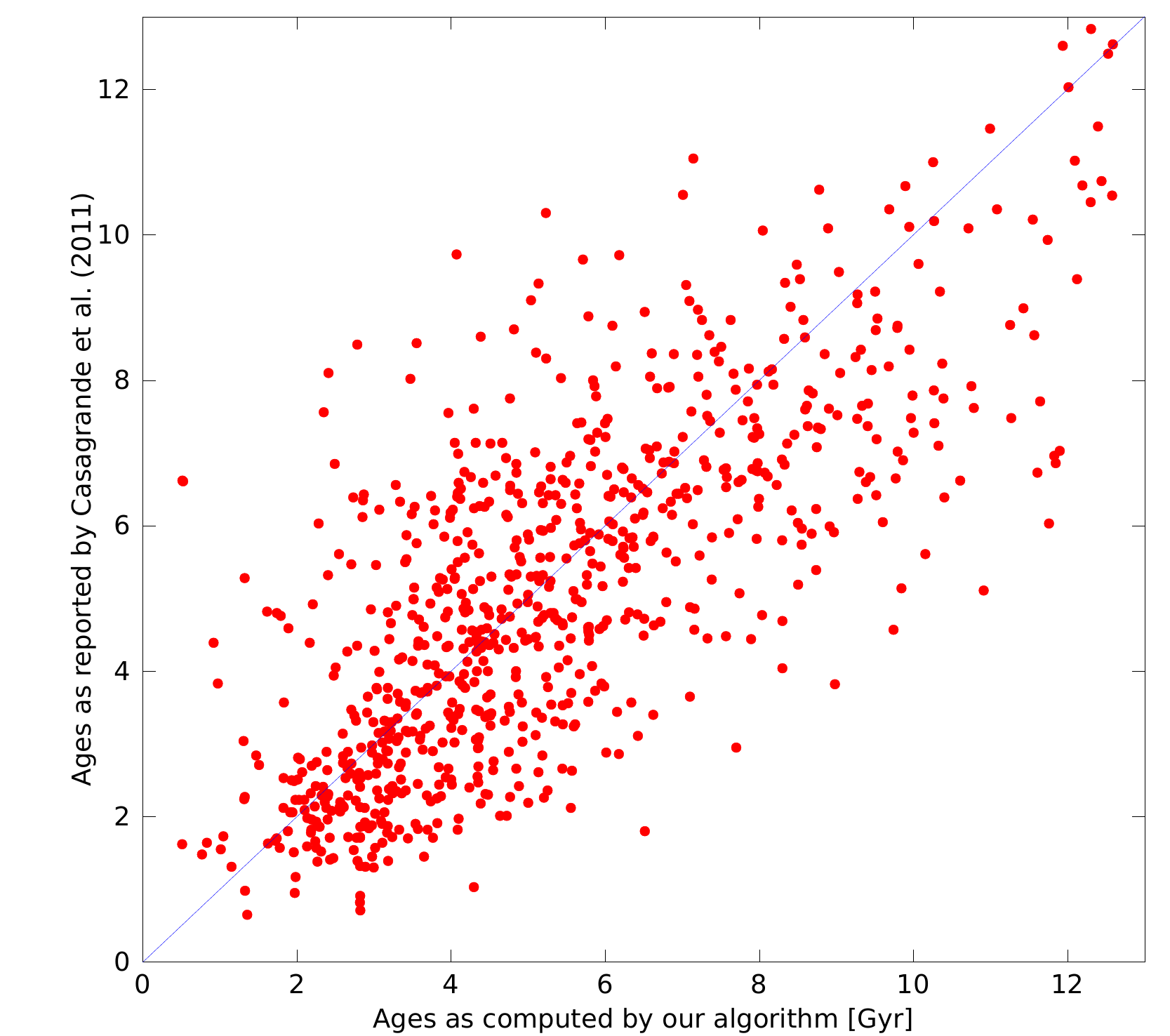}
 \caption{Age comparison between Casagrande estimation and ours (818 stars). The dispersion between the two measurements can be quantified through the linear correlation coefficient, that is $\sim0.75$.}
 \label{fig:CfrCasaAges}
\end{figure}

Our second step is to investigate whether exoplanet hosts are peculiar with respect to field stars not harbouring planets. The exoplanet hosts known so far are late spectral type stars located in the solar neighbourhood: $\sim90\%$ of the planet-hosting stars we analysed are closer than 200 pc. The stars contained in the SN catalogue represent an interesting comparison test because they represent a numerous sample of late spectral type MS-stars occupying almost the same volume as the exoplanet hosts. The age values reported by \citet{casagrande11} (C11 from here on) are computed following their own method, so we checked whether their results were consistent following the VF-SN subsample, which contains all the input parameters needed by our algorithm. We managed to obtain the age for 818 stars. We recall that the cut in spectral type is imposed by both C11 and \citet{valenti05} limits, and includes basically F, G and K stars. The plot of the expectation age reported by C11 versus our age 
values is shown in Fig. \ref{fig:CfrCasaAges}. The scatter of points around the bisector is expected given the high age uncertainties, and, in any case, good statistical agreement characterizes the two determinations. The median age for C11 values is $\sim4.9$ Gyr, which is very similar to our median age ($\sim4.8$ Gyr) for the common sample. This agreement between the two age determinations suggests that any comparison between C11 ages and ours is consistent. In addition, considering the median age value coming from the VF-SN sample, it is not surprising that the age is similar to the ages found for the samples of stars with planets analysed above. In fact, we obtained the VF-SN catalogue by cross-matching the SN sample with the catalogue of stars reported by \citet{valenti05}. 
The authors performed high-precision spectroscopy on stars taken from Keck, Lick and AAT planet search programme, thus, their stars present the typical selection effects characterizing stars with planets. Actually the median value we obtained for the VF-SN sample is the same as the SH value.

\begin{figure}
 \centering
 \includegraphics[width=\columnwidth]{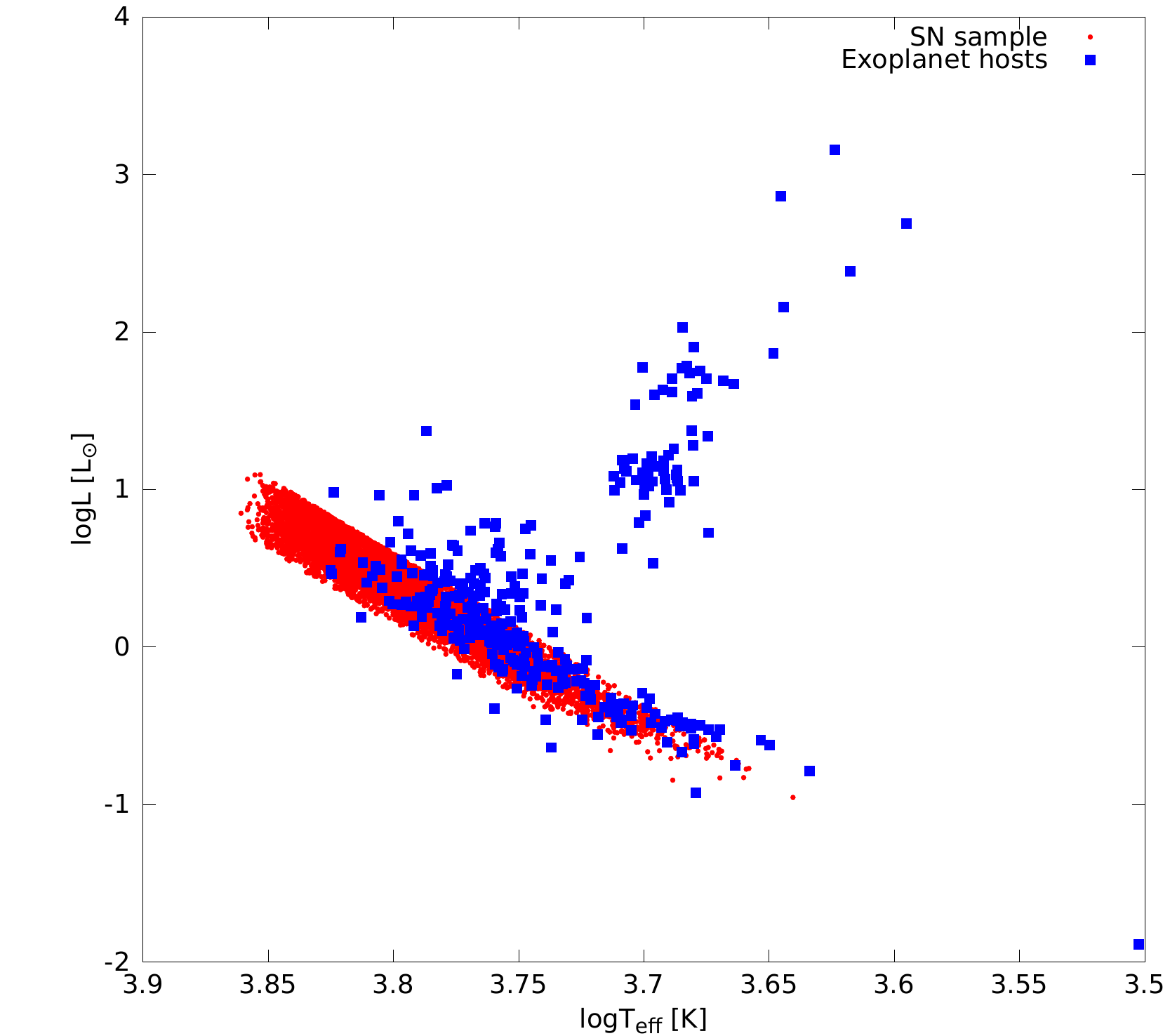}
 \caption{SN sample on the HRD (red dots). The very straight boundaries of the SN sample is simply a consequence of our identification of the MS through a strip as described in \S~\ref{subsec:SNcat}. All the planet-hosting stars we analysed are superimposed (blue squares).}
 \label{fig:CasaSweetHRD}
\end{figure}

As C11 age values suggest, the age distribution of all the 7044 stars belonging to the SN catalogue peaks in the [1.5, 2) Gyr bin with a median age value of $\sim2.6$ Gyr. This raw analysis may suggest that field stars are globally younger than planet-hosting stars. Instead, this comparison hides a bias, in fact, the SN catalogue contains a huge number of hot F-type stars with respect to planet-hosting stars, as shown in Fig. \ref{fig:CasaSweetHRD} (stars with $\log{\teff}\gtrsim3.8$). The earlier the spectral type, the faster the evolution of a star, thus, F-type stars are expected to be statistically younger than later spectral type stars.


\begin{figure}
 \centering
 \includegraphics[width=\columnwidth]{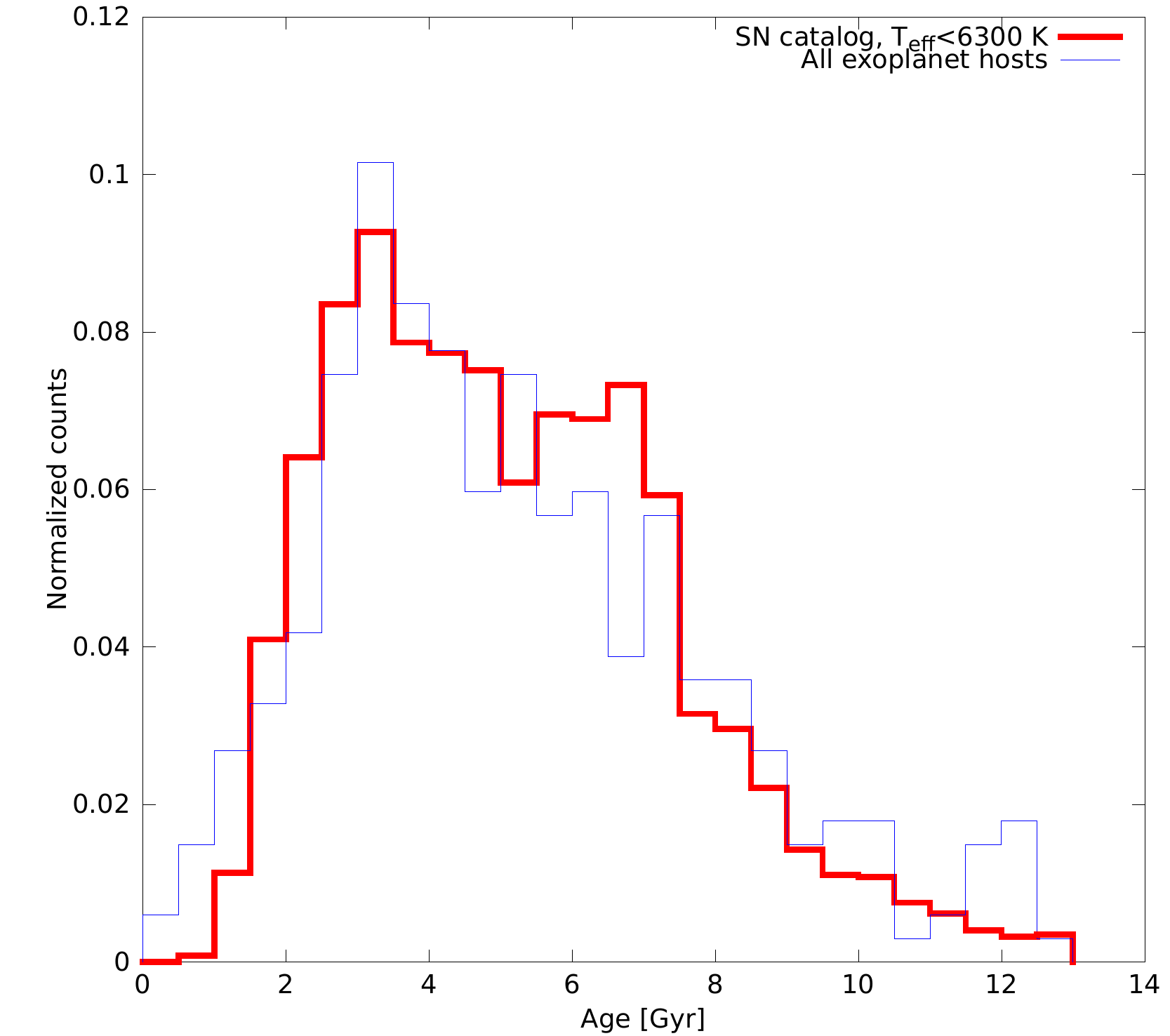}
 \caption{Age distribution of the MS stars colder than 6300 K belonging to the RSN catalogue (red stairs, 3713 stars) compared with the age distribution of all the planet-hosting stars we analysed (blue stairs, 335 stars).}
 \label{fig:CfrCasa6300Sweet}
\end{figure}

Analysing Fig. \ref{fig:CasaSweetHRD}, all the stars with $T_{\mathrm{eff}}>6300$ have been removed from the SN catalogue, remaining with 3713 stars (RSN catalogue). In this way, the comparison between the solar neighbourhood stars and the stars with planets is homogeneous in spectral type. The result is given by the age distributions in Fig. \ref{fig:CfrCasa6300Sweet}. They are almost consistent (KS p-value is 0.2): SN age distribution is less peaked, but they both have $4.8$ Gyr as median value. 

Two considerations should be added.
\begin{enumerate}
 \item The median age value of the stars with planets is very similar to the solar age value, but very different ($\sim1$ Gyr) from the histogram peak because of the extended tail towards old ages. According to their metallicities, there is no evidence that planet-hosting stars with $t\gtrsim9$ Gyr have population II properties. If we exclude this contamination, we may argue whether such a skew distribution may be due to a real distribution reflecting a prolonged star formation history \citep{rochaPinto00} or to an asymmetric propagation of errors.
 \item Assuming a solar age of $\sim4.6$ Gyr \citep{chaussidon07}, an age of $t=4.6$-4.8 Gyr appears older than that currently assumed for most of the thin disc population, where planet-hosting stars and the RSN sample are located. In fact, as summarized by \citet{prieto10}, \citet{reddy06} say that thin disc stars span a range between 1 and 9 Gyr, with the majority of them younger than 5 Gyr. \citet{holmberg09} and \citet{haywood08} set an older upper limit for the thin disc ages, however they both agree that most of these stars are younger than 4-5 Gyr. \citet{rochaPinto00}, using a different approach based on the stellar activity as age indicator, found three different peaks in the local star formation history, with the highest at very young ages below 1 Gyr.
 We caution that we limited our sample primarily to G-type stars. The other point is that we are sampling a very limited inter-arms volume (essentially $<200$ pc), as most of the recent studies based on single star age measurements. The literature does not present detailed studies of the ages of single disc stars far away from the solar neighbourhood. Thus, this lack of information does not allow us to perform a complete comparison between the evolutionary properties of our samples and those of the entire galactic disc. The extension of the stellar analysis to a distance larger than 200 pc would include younger active star-forming regions, such as the Orion Nebula or Taurus-Auriga complex. As a consequence, in deeper surveys we expect to include significantly younger stars.
\end{enumerate}

\section{Conclusions}\label{sec:conclusion}
We analysed a sample of 61 transiting-planet hosts to compute their ages and their peculiar parameters according to Padova isochrones. A priori, one could say that our particular sample is expected not to be affected by some typical biases that characterize those planet-hosting stars coming from radial velocity surveys. Spectroscopic targets are often deliberately chosen to be slow rotators and typically inactive. Instead, very high-precision photometry only requires bright stars in the solar neighbourhood for an adequate signal-to-noise ratio. Actually, once a possible transiting planet has been detected, the confirmation process involves spectroscopic analysis. Therefore, similar biases are expected in both spectroscopic and photometric surveys.

We found that the median age of our TPH sample is $\sim5$ Gyr. The subsample of TPH brighter than $V=12$ yields a median age of $\sim4.2$ Gyr. This slightly lower value is expected since brighter stars are on average younger. In order to comment the age distribution of TPH, we also considered 274 planet-hosting stars, whose planets have been detected though radial velocity method. Their age distribution peaks in the [3, 3.5) Gyr bin and it is synthesized by a median value of $\sim4.8$ Gyr. These three samples of stars are consistent from an evolutionary point of view. Slight differences are due to the paucity of stars belonging to the TPH catalogue and, in fact, a KS test does not suggest that TPH and SH come from a different distribution. Thus, spectroscopic and photometric targets are characterized by almost the same selection effects, and these biases bring the median of their age distribution around the solar age value.

In the second part we checked whether planet-hosting stars have peculiar ages with respect to field stars without planets of the solar neighbourhood. In case of a homogeneous comparison in terms of spectral type, solar neighborhood stars belonging to the RSN catalog have an age distribution very similar to that deriving from the all exoplanet hosts considered in this paper and the median age is $\sim4.8$ in both cases. With its age of 4.567 Gyr (as reported by \citealt{chaussidon07}), the Sun appears not to be a peculiar star, if compared with both the planet-hosting stars and the SN stars, whose spectral types span essentially from late-F to early K. However, it looks that we are sampling a limited inter-arms region, possibly older than the average thin disc population.

\begin{acknowledgements}
We thank the anonymous referee for the interesting and fruitful comments that improved our paper.
A. B. and S.O. acknowledge ``contratto ASI-INAF n.2015-019-R0 del 29 luglio 2015'' and support from INAF through the ``Progetti Premiali'' funding scheme of the Italian Ministry of Education, University, and Research. S. O. acknowledge financial support from University of Padova, as well. V. N. acknowledges support from INAF-OAPd through the grant ``Studio preparatorio per le osservazioni della missione ESA/CHEOPS'' (\#42/2013).
This research has made use of the Exoplanet Orbit Database and the Exoplanet Data Explorer at exoplanets.org. Moreover it has made use of SWEET-Cat: a catalog of stellar parameters for stars with planets at astro.up.pt (Centro de Astrofisica da Universidade do Porto).
\end{acknowledgements}

\onllongtab{
%
\begin{longtab}
\scriptsize
\begin{longtable}{lrrrrrrrrrrrrrr}
\caption{\label{tab:exostars} Planet-hosting stars parameters determined through Padova isochrones.}\\
\hline\hline
Star & $t$ & $\Delta t$ & $T_{\mathrm{eff}}$ & $\Delta T_{\mathrm{eff}}$ & $L$ & $\Delta L$ & $M$ & $\Delta M$ & $\log{g}$ & $\Delta\log{g}$ & $R$ & $\Delta R$ & Bin & Tr \\
     & (Gyr) & (Gyr) & (K) & (K) & ($L_{\odot}$) & ($L_{\odot}$) & ($M_{\odot}$) & ($M_{\odot}$) & (cm/s$^2$) & (cm/s$^2$) & ($R_{\odot}$) & ($R_{\odot}$) & & \\
\hline
\endfirsthead
\caption{continued.}\\
\hline\hline
Star & $t$ & $\Delta t$ & $T_{\mathrm{eff}}$ & $\Delta T_{\mathrm{eff}}$ & $L$ & $\Delta L$ & $M$ & $\Delta M$ & $\log{g}$ & $\Delta\log{g}$ & $R$ & $\Delta R$ & Bin & Tr \\
     & (Gyr) & (Gyr) & (K) & (K) & ($L_{\odot}$) & ($L_{\odot}$) & ($M_{\odot}$) & ($M_{\odot}$) & (cm/s$^2$) & (cm/s$^2$) & ($R_{\odot}$) & ($R_{\odot}$) & & \\
\hline
\endhead
\hline
\endfoot

  11 Com & 3.0 & 0.3 & 4741 & 12 & 109 & 1 & 1.37 & 0.04 & 2.19 & 0.02 & 15.5 & 0.2 & 1 & 0\\
  11 UMi & 4.1 & 1.0 & 4140 & 28 & 243 & 6 & 1.4 & 0.1 & 1.6 & 0.04 & 30.4 & 0.8 & 0 & 0\\
  14 And & 5.6 & 3.0 & 4775 & 49 & 56 & 1 & 1.2 & 0.2 & 2.4 & 0.1 & 11.0 & 0.3 & 0 & 0\\
  14 Her & 4.6 & 1.5 & 5313 & 18 & 0.61 & 0.01 & 0.97 & 0.01 & 4.48 & 0.02 & 0.93 & 0.01 & 0 & 0\\
  16 Cyg B & 5.2 & 1.0 & 5837 & 32 & 1.26 & 0.01 & 1.04 & 0.02 & 4.37 & 0.02 & 1.1 & 0.02 & 1 & 0\\
  18 Del & 1.0 & 0.1 & 5037 & 57 & 35 & 1 & 2.1 & 0.1 & 2.98 & 0.04 & 7.8 & 0.3 & 0 & 0\\
  24 Sex & 2.9 & 0.2 & 4917 & 10 & 16.4 & 0.1 & 1.49 & 0.02 & 3.11 & 0.01 & 5.6 & 0.04 & 0 & 0\\
  30 Ari B & 0.4 & 0.4 & 6313 & 24 & 1.98 & 0.03 & 1.25 & 0.02 & 4.38 & 0.01 & 1.18 & 0.02 & 1 & 0\\
  42 Dra & 6.5 & 1.7 & 4414 & 40 & 145 & 3 & 1.1 & 0.1 & 1.8 & 0.1 & 20.6 & 0.6 & 0 & 0\\
  51 Peg & 3.8 & 1.1 & 5857 & 39 & 1.34 & 0.03 & 1.09 & 0.02 & 4.37 & 0.02 & 1.13 & 0.03 & 0 & 0\\
  55 Cnc & 5.1 & 2.7 & 5310 & 32 & 0.59 & 0.01 & 0.95 & 0.02 & 4.49 & 0.03 & 0.91 & 0.02 & 1 & 1\\
  6 Lyn & 3.1 & 0.4 & 4973 & 28 & 14.8 & 0.1 & 1.4 & 0.1 & 3.15 & 0.03 & 5.2 & 0.1 & 0 & 0\\
  61 Vir & 5.2 & 2.1 & 5651 & 36 & 0.81 & 0.01 & 0.96 & 0.02 & 4.46 & 0.03 & 0.94 & 0.02 & 0 & 0\\
  7 CMa & 4.9 & 1.0 & 4782 & 39 & 11.3 & 0.2 & 1.3 & 0.1 & 3.17 & 0.04 & 4.9 & 0.1 & 0 & 0\\
  70 Vir & 8.1 & 0.4 & 5560 & 31 & 3.0 & 0.1 & 1.07 & 0.01 & 3.92 & 0.02 & 1.9 & 0.1 & 0 & 0\\
  75 Cet & 3.3 & 0.7 & 4742 & 14 & 53.7 & 0.3 & 1.4 & 0.1 & 2.52 & 0.03 & 10.9 & 0.1 & 0 & 0\\
  81 Cet & 3.2 & 0.5 & 4799 & 24 & 59.8 & 0.4 & 1.4 & 0.1 & 2.49 & 0.03 & 11.2 & 0.2 & 0 & 0\\
  91 Aqr & 8.1 & 2.8 & 4656 & 26 & 51 & 1 & 1.1 & 0.2 & 2.4 & 0.1 & 11.0 & 0.2 & 0 & 0\\
  BD +14 4559 & 6.9 & 4.2 & 4948 & 25 & 0.32 & 0.01 & 0.82 & 0.02 & 4.57 & 0.03 & 0.78 & 0.02 & 0 & 0\\
  BD +20 2457 & 10.5 & 1.7 & 4504 & 5 & 172 & 12 & 0.89 & 0.05 & 1.71 & 0.01 & 21.6 & 0.8 & 0 & 0\\
  CoRoT-18 & 7.5 & 4.5 & 5444 & 38 & 0.58 & 0.04 & 0.87 & 0.03 & 4.5 & 0.05 & 0.86 & 0.04 & 0 & 1\\
  CoRoT-19 & 5.1 & 0.8 & 6133 & 58 & 3.0 & 0.3 & 1.2 & 0.1 & 4.12 & 0.01 & 1.5 & 0.1 & 0 & 1\\
  CoRoT-7 & 4.5 & 4.3 & 5357 & 22 & 0.52 & 0.05 & 0.89 & 0.03 & 4.54 & 0.05 & 0.84 & 0.05 & 0 & 1\\
  HAT-P-11 & 5.2 & 3.1 & 4785 & 12 & 0.262 & 0.004 & 0.8 & 0.01 & 4.59 & 0.02 & 0.75 & 0.01 & 0 & 1\\
  HAT-P-13 & 5.6 & 0.9 & 5733 & 42 & 2.7 & 0.2 & 1.2 & 0.1 & 4.071 & 0.005 & 1.7 & 0.1 & 0 & 1\\
  HAT-P-14 & 1.1 & 0.8 & 6694 & 73 & 4.3 & 0.3 & 1.42 & 0.04 & 4.21 & 0.03 & 1.54 & 0.08 & 0 & 1\\
  HAT-P-17 & 8.9 & 2.9 & 5338 & 30 & 0.57 & 0.01 & 0.87 & 0.02 & 4.48 & 0.03 & 0.88 & 0.02 & 0 & 1\\
  HAT-P-22 & 12.3 & 0.6 & 5338 & 14 & 0.75 & 0.01 & 0.9 & 0.01 & 4.37 & 0.01 & 1.02 & 0.01 & 0 & 1\\
  HAT-P-24 & 2.2 & 1.2 & 6448 & 55 & 2.4 & 0.2 & 1.17 & 0.03 & 4.30 & 0.02 & 1.26 & 0.06 & 0 & 1\\
  HAT-P-26 & 3.6 & 0.3 & 5387 & 6 & 0.53 & 0.01 & 0.9 & 0.002 & 4.541 & 0.004 & 0.84 & 0.01 & 0 & 1\\
  HAT-P-27 & 7.9 & 2.1 & 5204 & 28 & 0.52 & 0.01 & 0.9 & 0.02 & 4.49 & 0.02 & 0.89 & 0.01 & 0 & 1\\
  HAT-P-36 & 10.0 & 0.7 & 5422 & 23 & 0.83 & 0.01 & 0.94 & 0.01 & 4.37 & 0.01 & 1.04 & 0.02 & 0 & 1\\
  HAT-P-4 & 5.4 & 0.9 & 5809 & 45 & 2.2 & 0.1 & 1.19 & 0.05 & 4.169 & 0.003 & 1.5 & 0.1 & 0 & 1\\
  HAT-P-7 & 0.8 & 0.7 & 6707 & 53 & 5.9 & 0.7 & 1.57 & 0.04 & 4.12 & 0.06 & 1.8 & 0.1 & 0 & 1\\
  HAT-P-9 & 1.8 & 0.4 & 6386 & 38 & 2.60 & 0.04 & 1.27 & 0.01 & 4.29 & 0.01 & 1.32 & 0.03 & 0 & 1\\
  HATS-1 & 7.5 & 1.1 & 5705 & 32 & 0.95 & 0.01 & 0.94 & 0.01 & 4.4 & 0.01 & 1.00 & 0.02 & 0 & 1\\
  HD 100655 & 9.4 & 2.1 & 4675 & 4 & 47 & 2 & 1.0 & 0.1 & 2.39 & 0.02 & 10.4 & 0.2 & 0 & 0\\
  HD 100777 & 5.9 & 1.8 & 5579 & 37 & 0.92 & 0.02 & 1.00 & 0.02 & 4.41 & 0.03 & 1.03 & 0.03 & 0 & 0\\
  HD 10180 & 4.5 & 1.1 & 5939 & 43 & 1.49 & 0.03 & 1.08 & 0.02 & 4.34 & 0.03 & 1.16 & 0.03 & 0 & 0\\
  HD 101930 & 5.7 & 4.3 & 5144 & 27 & 0.42 & 0.02 & 0.87 & 0.03 & 4.54 & 0.04 & 0.82 & 0.02 & 1 & 0\\
  HD 102117 & 6.1 & 1.2 & 5727 & 47 & 1.4 & 0.04 & 1.07 & 0.02 & 4.3 & 0.03 & 1.21 & 0.04 & 0 & 0\\
  HD 102195 & 5.9 & 3.5 & 5283 & 29 & 0.49 & 0.01 & 0.88 & 0.03 & 4.53 & 0.03 & 0.84 & 0.02 & 0 & 0\\
  HD 102272 & 11.6 & 1.2 & 4794 & 18 & 21.5 & 0.2 & 0.92 & 0.03 & 2.74 & 0.02 & 6.7 & 0.1 & 0 & 0\\
  HD 102329 & 4.1 & 0.8 & 4786 & 39 & 19.0 & 0.3 & 1.4 & 0.1 & 2.96 & 0.04 & 6.4 & 0.2 & 0 & 0\\
  HD 102365 & 10.8 & 2.0 & 5687 & 39 & 0.81 & 0.01 & 0.84 & 0.02 & 4.42 & 0.03 & 0.93 & 0.02 & 0 & 0\\
  HD 102956 & 2.9 & 0.2 & 4892 & 14 & 11.9 & 0.1 & 1.54 & 0.02 & 3.25 & 0.01 & 4.8 & 0.1 & 0 & 0\\
  HD 103197 & 3.4 & 2.0 & 5237 & 18 & 0.48 & 0.01 & 0.91 & 0.02 & 4.54 & 0.02 & 0.84 & 0.01 & 0 & 0\\
  HD 103774 & 2.0 & 0.1 & 6391 & 27 & 3.7 & 0.1 & 1.38 & 0.01 & 4.183 & 0.002 & 1.56 & 0.03 & 0 & 0\\
  HD 104067 & 9.0 & 3.4 & 4961 & 12 & 0.31 & 0.01 & 0.78 & 0.01 & 4.57 & 0.02 & 0.75 & 0.02 & 0 & 0\\
  HD 104985 & 4.9 & 1.2 & 4730 & 41 & 51 & 1 & 1.2 & 0.1 & 2.5 & 0.1 & 10.6 & 0.3 & 0 & 0\\
  HD 106252 & 6.4 & 1.4 & 5881 & 47 & 1.37 & 0.04 & 1.02 & 0.02 & 4.33 & 0.03 & 1.13 & 0.03 & 0 & 0\\
  HD 106270 & 3.8 & 0.2 & 5570 & 123 & 5.9 & 0.3 & 1.37 & 0.03 & 3.74 & 0.05 & 2.6 & 0.2 & 0 & 0\\
  HD 10647 & 1.8 & 0.9 & 6159 & 39 & 1.56 & 0.02 & 1.12 & 0.02 & 4.4 & 0.02 & 1.1 & 0.02 & 0 & 0\\
  HD 10697 & 7.5 & 0.4 & 5674 & 93 & 2.8 & 0.04 & 1.12 & 0.01 & 4.00 & 0.03 & 1.7 & 0.1 & 0 & 0\\
  HD 107148 & 4.0 & 1.0 & 5789 & 36 & 1.34 & 0.05 & 1.1 & 0.01 & 4.35 & 0.03 & 1.15 & 0.03 & 0 & 0\\
  HD 108147 & 1.3 & 0.5 & 6211 & 21 & 1.88 & 0.02 & 1.21 & 0.01 & 4.37 & 0.01 & 1.19 & 0.01 & 0 & 0\\
  HD 108863 & 3.3 & 0.5 & 4876 & 29 & 16.5 & 0.3 & 1.4 & 0.1 & 3.08 & 0.03 & 5.7 & 0.1 & 0 & 0\\
  HD 108874 & 6.3 & 0.7 & 5647 & 19 & 1.06 & 0.01 & 1.02 & 0.01 & 4.37 & 0.01 & 1.08 & 0.01 & 0 & 0\\
  HD 109246 & 2.5 & 0.8 & 5887 & 19 & 1.15 & 0.01 & 1.07 & 0.01 & 4.43 & 0.01 & 1.03 & 0.01 & 0 & 0\\
  HD 109749 & 4.1 & 0.7 & 5860 & 39 & 1.55 & 0.02 & 1.14 & 0.01 & 4.32 & 0.02 & 1.21 & 0.02 & 1 & 0\\
  HD 111232 & 11.6 & 1.5 & 5650 & 22 & 0.67 & 0.01 & 0.79 & 0.01 & 4.47 & 0.02 & 0.85 & 0.01 & 0 & 0\\
  HD 113337 & 1.5 & 0.9 & 6622 & 71 & 4.18 & 0.08 & 1.38 & 0.03 & 4.19 & 0.03 & 1.56 & 0.05 & 0 & 0\\
  HD 114386 & 8.8 & 2.8 & 4926 & 13 & 0.28 & 0.01 & 0.76 & 0.01 & 4.58 & 0.02 & 0.73 & 0.01 & 0 & 0\\
  HD 114613 & 5.0 & 0.1 & 5756 & 13 & 4.09 & 0.03 & 1.25 & 0.01 & 3.91 & 0.01 & 2.04 & 0.02 & 0 & 0\\
  HD 114729 & 9.3 & 0.6 & 5939 & 58 & 2.33 & 0.02 & 0.97 & 0.01 & 4.1 & 0.02 & 1.44 & 0.03 & 1 & 0\\
  HD 114762 & 12.4 & 0.6 & 6043 & 22 & 1.52 & 0.01 & 0.82 & 0.01 & 4.24 & 0.01 & 1.13 & 0.01 & 1 & 0\\
  HD 114783 & 10.4 & 2.4 & 5089 & 17 & 0.4 & 0.01 & 0.81 & 0.01 & 4.53 & 0.02 & 0.81 & 0.01 & 0 & 0\\
  HD 11506 & 2.0 & 0.6 & 6119 & 45 & 2.1 & 0.1 & 1.25 & 0.01 & 4.31 & 0.02 & 1.29 & 0.04 & 0 & 0\\
  HD 116029 & 4.9 & 1.1 & 4849 & 39 & 11.4 & 0.1 & 1.3 & 0.1 & 3.19 & 0.04 & 4.8 & 0.1 & 0 & 0\\
  HD 117207 & 6.6 & 1.0 & 5681 & 33 & 1.19 & 0.02 & 1.03 & 0.01 & 4.34 & 0.02 & 1.13 & 0.02 & 0 & 0\\
  HD 117618 & 4.0 & 1.3 & 6019 & 50 & 1.6 & 0.1 & 1.1 & 0.02 & 4.34 & 0.03 & 1.17 & 0.04 & 0 & 0\\
  HD 118203 & 5.4 & 0.5 & 5741 & 35 & 3.8 & 0.3 & 1.23 & 0.03 & 3.93 & 0.02 & 2.0 & 0.1 & 0 & 0\\
  HD 11964 & 8.5 & 0.5 & 5371 & 43 & 2.6 & 0.1 & 1.09 & 0.02 & 3.92 & 0.02 & 1.9 & 0.1 & 1 & 0\\
  HD 11977 & 2.9 & 0.2 & 4851 & 9 & 62.1 & 0.2 & 1.46 & 0.03 & 2.5 & 0.01 & 11.2 & 0.1 & 0 & 0\\
  HD 120084 & 4.4 & 1.2 & 4675 & 11 & 52 & 1 & 1.3 & 0.1 & 2.48 & 0.03 & 11.0 & 0.2 & 0 & 0\\
  HD 121504 & 1.9 & 1.0 & 6089 & 47 & 1.62 & 0.04 & 1.16 & 0.02 & 4.38 & 0.03 & 1.15 & 0.03 & 0 & 0\\
  HD 125595 & 8.0 & 3.7 & 4654 & 22 & 0.21 & 0.01 & 0.74 & 0.01 & 4.61 & 0.02 & 0.71 & 0.02 & 0 & 0\\
  HD 125612 & 3.1 & 0.3 & 5818 & 13 & 1.205 & 0.003 & 1.09 & 0.01 & 4.4 & 0.01 & 1.08 & 0.01 & 1 & 0\\
  HD 12661 & 3.3 & 0.6 & 5714 & 22 & 1.13 & 0.01 & 1.09 & 0.01 & 4.4 & 0.01 & 1.08 & 0.01 & 0 & 0\\
  HD 128311 & 8.5 & 3.6 & 4922 & 26 & 0.29 & 0.01 & 0.77 & 0.02 & 4.58 & 0.02 & 0.74 & 0.02 & 0 & 0\\
  HD 130322 & 6.1 & 2.9 & 5410 & 30 & 0.56 & 0.01 & 0.88 & 0.02 & 4.52 & 0.03 & 0.85 & 0.02 & 0 & 0\\
  HD 131496 & 4.0 & 0.7 & 4838 & 43 & 9.9 & 0.2 & 1.4 & 0.1 & 3.27 & 0.04 & 4.5 & 0.1 & 0 & 0\\
  HD 134987 & 5.4 & 0.5 & 5797 & 23 & 1.51 & 0.01 & 1.09 & 0.01 & 4.3 & 0.01 & 1.22 & 0.01 & 0 & 0\\
  HD 136418 & 5.0 & 1.0 & 4997 & 40 & 6.9 & 0.1 & 1.2 & 0.1 & 3.43 & 0.04 & 3.5 & 0.1 & 0 & 0\\
  HD 137388 & 7.3 & 3.8 & 5183 & 25 & 0.46 & 0.02 & 0.87 & 0.03 & 4.52 & 0.03 & 0.85 & 0.02 & 1 & 0\\
  HD 13908 & 3.9 & 0.5 & 6212 & 38 & 4.0 & 0.1 & 1.28 & 0.04 & 4.06 & 0.02 & 1.74 & 0.04 & 0 & 0\\
  HD 13931 & 5.3 & 1.3 & 5902 & 52 & 1.48 & 0.03 & 1.07 & 0.02 & 4.33 & 0.03 & 1.17 & 0.03 & 0 & 0\\
  HD 139357 & 7.2 & 1.8 & 4454 & 39 & 73.5 & 1.3 & 1.1 & 0.1 & 2.2 & 0.1 & 14.4 & 0.4 & 0 & 0\\
  HD 141937 & 3.2 & 0.5 & 5837 & 14 & 1.13 & 0.01 & 1.06 & 0.01 & 4.42 & 0.01 & 1.04 & 0.01 & 0 & 0\\
  HD 142 & 2.8 & 0.5 & 6321 & 67 & 2.81 & 0.05 & 1.25 & 0.01 & 4.24 & 0.03 & 1.4 & 0.04 & 1 & 0\\
  HD 142245 & 3.1 & 0.3 & 4831 & 28 & 13.1 & 0.2 & 1.52 & 0.05 & 3.19 & 0.03 & 5.2 & 0.1 & 0 & 0\\
  HD 142415 & 1.6 & 0.6 & 5869 & 12 & 1.16 & 0.02 & 1.1 & 0.01 & 4.44 & 0.01 & 1.04 & 0.01 & 0 & 0\\
  HD 145377 & 2.9 & 1.2 & 5979 & 46 & 1.43 & 0.04 & 1.11 & 0.02 & 4.38 & 0.03 & 1.12 & 0.03 & 0 & 0\\
  HD 145457 & 2.8 & 0.6 & 4772 & 45 & 41 & 1 & 1.5 & 0.1 & 2.66 & 0.05 & 9.4 & 0.2 & 0 & 0\\
  HD 1461 & 4.0 & 0.7 & 5807 & 20 & 1.20 & 0.01 & 1.07 & 0.01 & 4.39 & 0.01 & 1.08 & 0.01 & 0 & 0\\
  HD 147018 & 7.3 & 2.0 & 5526 & 29 & 0.78 & 0.02 & 0.94 & 0.02 & 4.44 & 0.03 & 0.97 & 0.02 & 0 & 0\\
  HD 147513 & 3.4 & 0.7 & 5827 & 21 & 1.01 & 0.01 & 1.02 & 0.01 & 4.45 & 0.01 & 0.99 & 0.01 & 1 & 0\\
  HD 148156 & 1.2 & 0.5 & 6156 & 23 & 1.84 & 0.03 & 1.22 & 0.01 & 4.36 & 0.01 & 1.19 & 0.02 & 0 & 0\\
  HD 148427 & 4.3 & 0.6 & 4993 & 44 & 6.2 & 0.1 & 1.32 & 0.05 & 3.51 & 0.03 & 3.3 & 0.1 & 0 & 0\\
  HD 149026 & 2.9 & 0.3 & 6116 & 44 & 2.8 & 0.1 & 1.302 & 0.005 & 4.2 & 0.02 & 1.49 & 0.04 & 0 & 1\\
  HD 149143 & 4.8 & 0.8 & 5792 & 58 & 2.2 & 0.1 & 1.21 & 0.03 & 4.17 & 0.03 & 1.5 & 0.1 & 0 & 0\\
  HD 1502 & 3.0 & 0.3 & 5006 & 25 & 11.5 & 0.2 & 1.46 & 0.04 & 3.29 & 0.02 & 4.5 & 0.1 & 0 & 0\\
  HD 152581 & 7.2 & 2.0 & 4991 & 45 & 16.1 & 0.2 & 1.0 & 0.1 & 3.0 & 0.1 & 5.4 & 0.1 & 0 & 0\\
  HD 153950 & 4.3 & 0.8 & 6136 & 64 & 2.24 & 0.03 & 1.15 & 0.02 & 4.25 & 0.03 & 1.33 & 0.04 & 0 & 0\\
  HD 154345 & 4.1 & 1.2 & 5557 & 15 & 0.62 & 0.002 & 0.9 & 0.01 & 4.53 & 0.01 & 0.85 & 0.01 & 0 & 0\\
  HD 154672 & 7.1 & 0.8 & 5754 & 51 & 1.81 & 0.02 & 1.09 & 0.02 & 4.2 & 0.03 & 1.36 & 0.03 & 0 & 0\\
  HD 154857 & 5.8 & 0.5 & 5740 & 46 & 4.4 & 0.3 & 1.13 & 0.03 & 3.83 & 0.03 & 2.1 & 0.1 & 0 & 0\\
  HD 155358 & 1.9 & 4.5 & 5966 & 53 & 2.11 & 0.02 & 1.1 & 0.1 & 4.2 & 0.04 & 1.36 & 0.03 & 0 & 0\\
  HD 156279 & 7.4 & 2.2 & 5449 & 31 & 0.7 & 0.01 & 0.93 & 0.02 & 4.45 & 0.03 & 0.94 & 0.02 & 0 & 0\\
  HD 156411 & 4.5 & 0.3 & 5886 & 29 & 5.1 & 0.3 & 1.23 & 0.02 & 3.85 & 0.02 & 2.2 & 0.1 & 0 & 0\\
  HD 156668 & 10.2 & 2.8 & 4857 & 18 & 0.27 & 0.01 & 0.75 & 0.01 & 4.58 & 0.01 & 0.73 & 0.01 & 0 & 0\\
  HD 158038 & 3.2 & 0.4 & 4839 & 29 & 11.9 & 0.1 & 1.5 & 0.1 & 3.23 & 0.03 & 4.9 & 0.1 & 0 & 0\\
  HD 159243 & 2.0 & 0.3 & 6071 & 19 & 1.45 & 0.05 & 1.12 & 0.01 & 4.4 & 0.01 & 1.09 & 0.03 & 0 & 0\\
  HD 159868 & 6.3 & 0.5 & 5583 & 54 & 3.8 & 0.2 & 1.13 & 0.03 & 3.85 & 0.03 & 2.1 & 0.1 & 0 & 0\\
  HD 16141 & 6.5 & 0.6 & 5856 & 60 & 2.46 & 0.03 & 1.13 & 0.02 & 4.12 & 0.03 & 1.53 & 0.04 & 1 & 0\\
  HD 16175 & 4.1 & 0.8 & 6009 & 44 & 3.35 & 0.02 & 1.3 & 0.05 & 4.09 & 0.02 & 1.69 & 0.03 & 0 & 0\\
  HD 162020 & 3.1 & 2.7 & 4807 & 17 & 0.22 & 0.01 & 0.75 & 0.01 & 4.63 & 0.01 & 0.68 & 0.01 & 0 & 0\\
  HD 163607 & 8.3 & 0.5 & 5508 & 15 & 2.6 & 0.1 & 1.1 & 0.02 & 3.98 & 0.01 & 1.8 & 0.1 & 0 & 0\\
  HD 16417 & 6.9 & 0.4 & 5818 & 51 & 2.74 & 0.01 & 1.12 & 0.01 & 4.06 & 0.02 & 1.63 & 0.03 & 0 & 0\\
  HD 164509 & 3.2 & 0.8 & 5860 & 31 & 1.31 & 0.02 & 1.1 & 0.01 & 4.38 & 0.02 & 1.11 & 0.02 & 0 & 0\\
  HD 164922 & 7.9 & 2.7 & 5439 & 38 & 0.7 & 0.01 & 0.93 & 0.02 & 4.45 & 0.03 & 0.95 & 0.02 & 0 & 0\\
  HD 167042 & 3.1 & 0.3 & 4989 & 32 & 10.7 & 0.1 & 1.46 & 0.05 & 3.31 & 0.03 & 4.4 & 0.1 & 0 & 0\\
  HD 168443 & 10.0 & 0.3 & 5646 & 36 & 2.08 & 0.01 & 1.02 & 0.01 & 4.08 & 0.01 & 1.51 & 0.02 & 0 & 0\\
  HD 168746 & 12.0 & 0.9 & 5637 & 26 & 1.04 & 0.01 & 0.9 & 0.01 & 4.32 & 0.01 & 1.07 & 0.01 & 0 & 0\\
  HD 169830 & 2.82 & 0.03 & 6276 & 12 & 4.656 & 0.003 & 1.3975 & 4.0E-4 & 4.052 & 0.004 & 1.83 & 0.01 & 0 & 0\\
  HD 170469 & 4.7 & 0.9 & 5866 & 54 & 1.7 & 0.1 & 1.14 & 0.01 & 4.28 & 0.03 & 1.27 & 0.05 & 0 & 0\\
  HD 171028 & 8.2 & 1.1 & 5771 & 46 & 3.9 & 0.5 & 0.98 & 0.04 & 3.84 & 0.03 & 2.0 & 0.2 & 0 & 0\\
  HD 171238 & 4.0 & 1.2 & 5570 & 21 & 0.774 & 0.003 & 0.99 & 0.01 & 4.47 & 0.01 & 0.95 & 0.01 & 0 & 0\\
  HD 17156 & 4.9 & 0.6 & 5943 & 38 & 2.46 & 0.05 & 1.2 & 0.03 & 4.17 & 0.02 & 1.48 & 0.03 & 0 & 1\\
  HD 173416 & 1.8 & 0.7 & 4790 & 37 & 80 & 2 & 1.8 & 0.2 & 2.5 & 0.1 & 13.0 & 0.3 & 0 & 0\\
  HD 175541 & 2.9 & 0.2 & 5093 & 23 & 10.0 & 0.1 & 1.45 & 0.03 & 3.37 & 0.02 & 4.07 & 0.05 & 0 & 0\\
  HD 177830 & 10.2 & 1.7 & 4735 & 31 & 5.3 & 0.1 & 1.1 & 0.1 & 3.39 & 0.04 & 3.4 & 0.1 & 1 & 0\\
  HD 178911 B & 4.8 & 1.3 & 5642 & 29 & 1.00 & 0.02 & 1.03 & 0.02 & 4.4 & 0.02 & 1.05 & 0.02 & 1 & 0\\
  HD 179079 & 7.8 & 0.4 & 5649 & 47 & 2.392 & 0.004 & 1.11 & 0.01 & 4.06 & 0.02 & 1.62 & 0.03 & 0 & 0\\
  HD 179949 & 1.2 & 0.6 & 6220 & 28 & 1.95 & 0.01 & 1.23 & 0.01 & 4.36 & 0.01 & 1.2 & 0.01 & 0 & 0\\
  HD 180314 & 0.9 & 0.2 & 4946 & 55 & 40 & 1 & 2.3 & 0.1 & 2.92 & 0.05 & 8.7 & 0.3 & 0 & 0\\
  HD 180902 & 3.3 & 0.5 & 5001 & 44 & 9.4 & 0.1 & 1.4 & 0.1 & 3.36 & 0.04 & 4.1 & 0.1 & 0 & 0\\
  HD 181342 & 3.6 & 0.6 & 4856 & 41 & 12.3 & 0.4 & 1.4 & 0.1 & 3.2 & 0.04 & 5.0 & 0.2 & 0 & 0\\
  HD 181433 & 7.4 & 3.4 & 4909 & 20 & 0.34 & 0.01 & 0.84 & 0.02 & 4.55 & 0.02 & 0.8 & 0.02 & 0 & 0\\
  HD 181720 & 12.4 & 0.5 & 5840 & 49 & 2.112 & 0.003 & 0.87 & 0.01 & 4.06 & 0.02 & 1.42 & 0.02 & 0 & 0\\
  HD 183263 & 4.5 & 0.8 & 5870 & 56 & 1.8 & 0.1 & 1.16 & 0.02 & 4.28 & 0.03 & 1.29 & 0.05 & 0 & 0\\
  HD 185269 & 4.1 & 0.5 & 6023 & 43 & 4.5 & 0.1 & 1.3 & 0.04 & 3.97 & 0.03 & 2.0 & 0.1 & 1 & 0\\
  HD 187085 & 2.7 & 0.8 & 6163 & 53 & 2.0 & 0.1 & 1.19 & 0.02 & 4.31 & 0.03 & 1.26 & 0.04 & 0 & 0\\
  HD 187123 & 5.6 & 1.3 & 5853 & 53 & 1.44 & 0.02 & 1.06 & 0.02 & 4.32 & 0.03 & 1.17 & 0.03 & 0 & 0\\
  HD 18742 & 3.9 & 0.8 & 4956 & 40 & 14.0 & 0.2 & 1.3 & 0.1 & 3.14 & 0.04 & 5.1 & 0.1 & 0 & 0\\
  HD 188015 & 5.9 & 1.3 & 5722 & 52 & 1.41 & 0.03 & 1.08 & 0.02 & 4.3 & 0.03 & 1.21 & 0.03 & 1 & 0\\
  HD 189733 & 5.3 & 3.8 & 5019 & 23 & 0.327 & 0.003 & 0.81 & 0.02 & 4.58 & 0.02 & 0.76 & 0.01 & 1 & 1\\
  HD 190360 & 7.3 & 1.6 & 5628 & 47 & 1.12 & 0.03 & 1.01 & 0.02 & 4.34 & 0.03 & 1.12 & 0.03 & 1 & 0\\
  HD 190647 & 8.7 & 0.4 & 5630 & 48 & 2.19 & 0.01 & 1.07 & 0.01 & 4.07 & 0.02 & 1.56 & 0.03 & 0 & 0\\
  HD 192263 & 5.9 & 3.9 & 4980 & 20 & 0.3 & 0.01 & 0.78 & 0.02 & 4.59 & 0.02 & 0.73 & 0.01 & 0 & 0\\
  HD 192310 & 8.1 & 3.2 & 5153 & 21 & 0.4 & 0.01 & 0.82 & 0.02 & 4.54 & 0.02 & 0.79 & 0.01 & 0 & 0\\
  HD 192699 & 3.1 & 0.4 & 5097 & 36 & 11.1 & 0.1 & 1.39 & 0.05 & 3.31 & 0.03 & 4.3 & 0.1 & 0 & 0\\
  HD 195019 & 7.7 & 0.7 & 5825 & 56 & 2.23 & 0.02 & 1.08 & 0.01 & 4.13 & 0.02 & 1.47 & 0.04 & 1 & 0\\
  HD 196050 & 5.4 & 0.7 & 5884 & 47 & 2.09 & 0.02 & 1.15 & 0.02 & 4.2 & 0.02 & 1.4 & 0.03 & 1 & 0\\
  HD 19994 & 3.1 & 0.3 & 6164 & 62 & 3.78 & 0.04 & 1.35 & 0.01 & 4.1 & 0.02 & 1.71 & 0.04 & 1 & 0\\
  HD 200964 & 3.1 & 0.4 & 5059 & 34 & 12.8 & 0.2 & 1.4 & 0.1 & 3.23 & 0.03 & 4.7 & 0.1 & 0 & 0\\
  HD 202206 & 2.9 & 1.0 & 5719 & 26 & 1.04 & 0.01 & 1.07 & 0.02 & 4.43 & 0.02 & 1.04 & 0.01 & 0 & 0\\
  HD 2039 & 4.4 & 0.9 & 5927 & 60 & 2.1 & 0.1 & 1.19 & 0.02 & 4.23 & 0.04 & 1.4 & 0.1 & 0 & 0\\
  HD 204313 & 4.3 & 1.8 & 5783 & 48 & 1.18 & 0.03 & 1.06 & 0.03 & 4.39 & 0.04 & 1.08 & 0.03 & 0 & 0\\
  HD 204941 & 3.9 & 3.3 & 5072 & 17 & 0.297 & 0.003 & 0.77 & 0.02 & 4.62 & 0.02 & 0.71 & 0.01 & 1 & 0\\
  HD 205739 & 2.9 & 0.2 & 6247 & 40 & 3.581 & 0.003 & 1.335 & 0.003 & 4.14 & 0.01 & 1.62 & 0.02 & 0 & 0\\
  HD 206610 & 3.0 & 0.3 & 4836 & 30 & 18 & 1 & 1.51 & 0.05 & 3.05 & 0.03 & 6.0 & 0.2 & 0 & 0\\
  HD 20782 & 5.4 & 1.3 & 5876 & 31 & 1.20 & 0.03 & 1.01 & 0.02 & 4.39 & 0.03 & 1.06 & 0.03 & 1 & 0\\
  HD 207832 & 1.4 & 0.8 & 5676 & 40 & 0.82 & 0.04 & 1.03 & 0.02 & 4.5 & 0.01 & 0.94 & 0.04 & 0 & 0\\
  HD 20794 & 11.6 & 1.5 & 5602 & 20 & 0.642 & 0.003 & 0.8 & 0.01 & 4.47 & 0.02 & 0.85 & 0.01 & 0 & 0\\
  HD 208487 & 2.3 & 0.9 & 6143 & 47 & 1.76 & 0.05 & 1.16 & 0.02 & 4.36 & 0.03 & 1.17 & 0.03 & 0 & 0\\
  HD 20868 & 8.2 & 2.7 & 4769 & 24 & 0.25 & 0.01 & 0.77 & 0.01 & 4.59 & 0.01 & 0.73 & 0.02 & 0 & 0\\
  HD 209458 & 4.4 & 1.2 & 6047 & 62 & 1.8 & 0.04 & 1.11 & 0.02 & 4.3 & 0.04 & 1.22 & 0.04 & 0 & 1\\
  HD 210277 & 8.8 & 1.9 & 5530 & 40 & 0.92 & 0.03 & 0.96 & 0.02 & 4.37 & 0.03 & 1.05 & 0.03 & 0 & 0\\
  HD 210702 & 3.1 & 0.3 & 4946 & 25 & 12.9 & 0.1 & 1.47 & 0.04 & 3.22 & 0.02 & 4.9 & 0.1 & 0 & 0\\
  HD 212771 & 2.9 & 0.1 & 5008 & 14 & 15.1 & 0.2 & 1.45 & 0.02 & 3.16 & 0.02 & 5.2 & 0.1 & 0 & 0\\
  HD 213240 & 4.6 & 0.6 & 6029 & 37 & 2.6 & 0.1 & 1.2 & 0.02 & 4.17 & 0.02 & 1.48 & 0.03 & 1 & 0\\
  HD 215497 & 9.9 & 2.8 & 5128 & 12 & 0.47 & 0.02 & 0.86 & 0.02 & 4.49 & 0.03 & 0.87 & 0.02 & 0 & 0\\
  HD 216437 & 5.2 & 0.7 & 5898 & 37 & 2.23 & 0.03 & 1.17 & 0.03 & 4.19 & 0.02 & 1.43 & 0.03 & 0 & 0\\
  HD 216770 & 5.4 & 2.9 & 5406 & 39 & 0.66 & 0.01 & 0.95 & 0.03 & 4.48 & 0.03 & 0.93 & 0.02 & 0 & 0\\
  HD 217107 & 4.2 & 1.0 & 5676 & 31 & 1.14 & 0.01 & 1.08 & 0.01 & 4.38 & 0.02 & 1.11 & 0.02 & 0 & 0\\
  HD 217786 & 6.8 & 0.9 & 6031 & 55 & 1.93 & 0.04 & 1.03 & 0.02 & 4.23 & 0.03 & 1.27 & 0.04 & 0 & 0\\
  HD 218566 & 8.0 & 3.1 & 4880 & 16 & 0.3 & 0.01 & 0.8 & 0.01 & 4.57 & 0.02 & 0.77 & 0.02 & 0 & 0\\
  HD 219828 & 5.2 & 0.8 & 5921 & 53 & 2.74 & 0.03 & 1.2 & 0.04 & 4.11 & 0.03 & 1.58 & 0.04 & 0 & 0\\
  HD 220773 & 6.3 & 0.1 & 5852 & 26 & 3.16 & 0.01 & 1.154 & 0.003 & 4.02 & 0.01 & 1.73 & 0.02 & 0 & 0\\
  HD 221287 & 2.8 & 0.3 & 6193 & 20 & 2.0 & 0.1 & 1.17 & 0.01 & 4.33 & 0.01 & 1.22 & 0.03 & 0 & 0\\
  HD 222155 & 8.1 & 0.4 & 5814 & 43 & 2.9 & 0.1 & 1.05 & 0.01 & 4.00 & 0.01 & 1.7 & 0.1 & 0 & 0\\
  HD 222582 & 6.2 & 1.1 & 5851 & 32 & 1.24 & 0.01 & 1.01 & 0.02 & 4.36 & 0.02 & 1.09 & 0.02 & 1 & 0\\
  HD 224693 & 3.9 & 0.5 & 5972 & 49 & 4.1 & 0.1 & 1.35 & 0.04 & 4.01 & 0.01 & 1.89 & 0.05 & 0 & 0\\
  HD 22781 & 7.5 & 2.9 & 5152 & 27 & 0.32 & 0.01 & 0.74 & 0.02 & 4.6 & 0.02 & 0.71 & 0.02 & 0 & 0\\
  HD 23079 & 4.1 & 1.4 & 6039 & 44 & 1.37 & 0.03 & 1.03 & 0.02 & 4.39 & 0.03 & 1.07 & 0.03 & 0 & 0\\
  HD 23127 & 4.4 & 0.6 & 5841 & 45 & 3.08 & 0.02 & 1.29 & 0.03 & 4.07 & 0.02 & 1.72 & 0.03 & 0 & 0\\
  HD 231701 & 3.7 & 0.5 & 6211 & 71 & 2.94 & 0.05 & 1.23 & 0.01 & 4.18 & 0.03 & 1.48 & 0.05 & 0 & 0\\
  HD 23596 & 4.0 & 0.7 & 5979 & 68 & 2.65 & 0.03 & 1.25 & 0.03 & 4.17 & 0.03 & 1.52 & 0.04 & 0 & 0\\
  HD 24040 & 4.9 & 0.9 & 5910 & 53 & 1.78 & 0.05 & 1.13 & 0.02 & 4.27 & 0.03 & 1.27 & 0.04 & 0 & 0\\
  HD 25171 & 4.8 & 0.9 & 6131 & 57 & 1.94 & 0.02 & 1.08 & 0.02 & 4.28 & 0.03 & 1.24 & 0.03 & 0 & 0\\
  HD 2638 & 5.1 & 4.1 & 5173 & 26 & 0.42 & 0.01 & 0.87 & 0.03 & 4.55 & 0.03 & 0.81 & 0.02 & 0 & 0\\
  HD 27894 & 6.9 & 4.3 & 4923 & 32 & 0.33 & 0.01 & 0.83 & 0.03 & 4.56 & 0.03 & 0.79 & 0.02 & 0 & 0\\
  HD 28185 & 5.5 & 4.2 & 5615 & 56 & 1.17 & 0.02 & 1.0 & 0.1 & 4.33 & 0.03 & 1.15 & 0.03 & 0 & 0\\
  HD 28254 & 7.8 & 0.4 & 5607 & 37 & 2.19 & 0.01 & 1.11 & 0.01 & 4.08 & 0.02 & 1.57 & 0.02 & 1 & 0\\
  HD 28678 & 6.1 & 1.7 & 4798 & 43 & 24.0 & 0.4 & 1.1 & 0.1 & 2.8 & 0.1 & 7.1 & 0.2 & 0 & 0\\
  HD 290327 & 11.5 & 1.3 & 5543 & 13 & 0.74 & 0.02 & 0.85 & 0.01 & 4.42 & 0.02 & 0.93 & 0.02 & 0 & 0\\
  HD 2952 & 3.1 & 0.3 & 4755 & 18 & 61.5 & 0.4 & 1.5 & 0.1 & 2.47 & 0.02 & 11.6 & 0.1 & 0 & 0\\
  HD 30177 & 5.9 & 1.1 & 5596 & 32 & 1.09 & 0.01 & 1.05 & 0.01 & 4.36 & 0.02 & 1.11 & 0.02 & 0 & 0\\
  HD 30562 & 3.7 & 0.5 & 6000 & 55 & 2.8 & 0.02 & 1.28 & 0.02 & 4.16 & 0.02 & 1.55 & 0.03 & 0 & 0\\
  HD 30856 & 7.3 & 1.8 & 4911 & 41 & 10.0 & 0.2 & 1.1 & 0.1 & 3.19 & 0.05 & 4.4 & 0.1 & 0 & 0\\
  HD 31253 & 4.0 & 0.7 & 6105 & 63 & 2.9 & 0.1 & 1.25 & 0.04 & 4.16 & 0.03 & 1.5 & 0.1 & 0 & 0\\
  HD 32518 & 5.8 & 1.5 & 4610 & 40 & 47 & 1 & 1.2 & 0.1 & 2.4 & 0.1 & 10.8 & 0.3 & 0 & 0\\
  HD 330075 & 6.1 & 4.0 & 5127 & 26 & 0.4 & 0.03 & 0.84 & 0.02 & 4.55 & 0.03 & 0.8 & 0.04 & 0 & 0\\
  HD 33142 & 3.3 & 0.4 & 4980 & 37 & 10.5 & 0.2 & 1.4 & 0.1 & 3.31 & 0.03 & 4.4 & 0.1 & 0 & 0\\
  HD 33283 & 3.9 & 0.6 & 5980 & 54 & 4.43 & 0.02 & 1.37 & 0.04 & 3.98 & 0.03 & 1.97 & 0.04 & 0 & 0\\
  HD 34445 & 3.7 & 0.6 & 6038 & 53 & 2.1 & 0.1 & 1.19 & 0.01 & 4.27 & 0.03 & 1.32 & 0.04 & 1 & 0\\
  HD 3651 & 6.9 & 2.8 & 5271 & 26 & 0.51 & 0.01 & 0.88 & 0.02 & 4.51 & 0.02 & 0.86 & 0.01 & 1 & 0\\
  HD 37124 & 11.1 & 1.7 & 5733 & 37 & 0.81 & 0.01 & 0.82 & 0.02 & 4.42 & 0.02 & 0.92 & 0.02 & 0 & 0\\
  HD 37605 & 4.2 & 1.4 & 5364 & 25 & 0.62 & 0.01 & 0.96 & 0.01 & 4.49 & 0.01 & 0.91 & 0.02 & 0 & 0\\
  HD 38283 & 6.5 & 0.6 & 6080 & 59 & 2.56 & 0.01 & 1.07 & 0.02 & 4.14 & 0.03 & 1.45 & 0.03 & 0 & 0\\
  HD 38529 & 3.98 & 0.03 & 5526 & 17 & 5.81 & 0.03 & 1.412 & 0.003 & 3.74 & 0.01 & 2.64 & 0.02 & 1 & 0\\
  HD 38801 & 4.8 & 0.3 & 5323 & 52 & 3.7 & 0.1 & 1.28 & 0.02 & 3.82 & 0.02 & 2.3 & 0.1 & 0 & 0\\
  HD 39091 & 2.8 & 0.8 & 6018 & 31 & 1.5 & 0.02 & 1.12 & 0.01 & 4.37 & 0.02 & 1.13 & 0.02 & 0 & 0\\
  HD 40307 & 6.9 & 4.0 & 4956 & 18 & 0.243 & 0.003 & 0.71 & 0.02 & 4.63 & 0.02 & 0.67 & 0.01 & 0 & 0\\
  HD 40979 & 1.5 & 0.6 & 6163 & 25 & 1.82 & 0.03 & 1.21 & 0.01 & 4.36 & 0.02 & 1.19 & 0.02 & 1 & 0\\
  HD 4113 & 5.8 & 1.6 & 5717 & 46 & 1.16 & 0.04 & 1.03 & 0.02 & 4.36 & 0.03 & 1.1 & 0.03 & 1 & 0\\
  HD 4203 & 7.3 & 0.9 & 5640 & 57 & 1.71 & 0.02 & 1.09 & 0.02 & 4.2 & 0.03 & 1.37 & 0.04 & 0 & 0\\
  HD 4208 & 7.4 & 2.4 & 5678 & 33 & 0.71 & 0.01 & 0.86 & 0.02 & 4.48 & 0.03 & 0.88 & 0.02 & 0 & 0\\
  HD 4308 & 0.4 & 2.2 & 5674 & 51 & 1.02 & 0.03 & 0.96 & 0.03 & 4.37 & 0.02 & 1.05 & 0.03 & 0 & 0\\
  HD 4313 & 3.0 & 0.3 & 4920 & 21 & 14.0 & 0.2 & 1.49 & 0.04 & 3.18 & 0.02 & 5.2 & 0.1 & 0 & 0\\
  HD 43197 & 4.4 & 2.1 & 5457 & 33 & 0.75 & 0.02 & 1.00 & 0.02 & 4.46 & 0.03 & 0.97 & 0.03 & 0 & 0\\
  HD 43691 & 2.1 & 1.7 & 6101 & 67 & 3.2 & 0.1 & 1.33 & 0.03 & 4.14 & 0.03 & 1.6 & 0.1 & 0 & 0\\
  HD 44219 & 9.7 & 0.8 & 5739 & 50 & 1.82 & 0.02 & 1.01 & 0.01 & 4.16 & 0.02 & 1.37 & 0.03 & 0 & 0\\
  HD 45350 & 7.1 & 0.9 & 5683 & 35 & 1.43 & 0.02 & 1.06 & 0.01 & 4.27 & 0.02 & 1.24 & 0.02 & 0 & 0\\
  HD 45364 & 5.8 & 2.4 & 5523 & 28 & 0.575 & 0.004 & 0.86 & 0.02 & 4.53 & 0.02 & 0.83 & 0.01 & 0 & 0\\
  HD 45652 & 5.4 & 2.7 & 5342 & 31 & 0.61 & 0.01 & 0.94 & 0.02 & 4.49 & 0.03 & 0.91 & 0.02 & 0 & 0\\
  HD 46375 & 11.9 & 1.1 & 5379 & 19 & 0.77 & 0.01 & 0.91 & 0.01 & 4.38 & 0.01 & 1.01 & 0.01 & 1 & 0\\
  HD 47186 & 5.3 & 0.8 & 5729 & 24 & 1.19 & 0.02 & 1.05 & 0.01 & 4.36 & 0.02 & 1.11 & 0.02 & 0 & 0\\
  HD 49674 & 3.6 & 0.8 & 5655 & 25 & 0.99 & 0.02 & 1.06 & 0.01 & 4.42 & 0.01 & 1.04 & 0.02 & 0 & 0\\
  HD 50499 & 2.3 & 0.4 & 6112 & 30 & 2.26 & 0.04 & 1.27 & 0.01 & 4.28 & 0.02 & 1.34 & 0.03 & 0 & 0\\
  HD 50554 & 2.1 & 0.5 & 6047 & 17 & 1.37 & 0.01 & 1.1 & 0.01 & 4.41 & 0.01 & 1.07 & 0.01 & 0 & 0\\
  HD 52265 & 2.2 & 0.7 & 6183 & 41 & 2.06 & 0.03 & 1.22 & 0.01 & 4.32 & 0.02 & 1.25 & 0.02 & 0 & 0\\
  HD 5319 & 6.1 & 1.4 & 4888 & 39 & 8.2 & 0.1 & 1.2 & 0.1 & 3.3 & 0.04 & 4.0 & 0.1 & 0 & 0\\
  HD 5608 & 3.0 & 0.3 & 4897 & 25 & 13.1 & 0.3 & 1.5 & 0.04 & 3.2 & 0.02 & 5.0 & 0.1 & 0 & 0\\
  HD 5891 & 5.7 & 1.5 & 4796 & 41 & 39.1 & 0.4 & 1.1 & 0.1 & 2.57 & 0.05 & 9.1 & 0.2 & 0 & 0\\
  HD 60532 & 3.0 & 0.2 & 6188 & 17 & 9.3 & 0.1 & 1.46 & 0.03 & 3.75 & 0.02 & 2.66 & 0.03 & 0 & 0\\
  HD 63454 & 2.4 & 3.1 & 4787 & 12 & 0.24 & 0.01 & 0.79 & 0.01 & 4.62 & 0.02 & 0.72 & 0.01 & 0 & 0\\
  HD 63765 & 7.9 & 3.1 & 5474 & 35 & 0.58 & 0.01 & 0.85 & 0.02 & 4.51 & 0.03 & 0.85 & 0.02 & 0 & 0\\
  HD 6434 & 12.2 & 0.7 & 5907 & 21 & 1.208 & 0.003 & 0.83 & 0.01 & 4.31 & 0.01 & 1.05 & 0.01 & 0 & 0\\
  HD 65216 & 4.6 & 3.1 & 5694 & 45 & 0.72 & 0.02 & 0.91 & 0.03 & 4.51 & 0.04 & 0.87 & 0.02 & 1 & 0\\
  HD 66428 & 5.9 & 0.8 & 5721 & 29 & 1.28 & 0.02 & 1.06 & 0.01 & 4.33 & 0.02 & 1.15 & 0.02 & 0 & 0\\
  HD 6718 & 6.2 & 2.0 & 5805 & 46 & 1.06 & 0.02 & 0.97 & 0.02 & 4.4 & 0.03 & 1.02 & 0.03 & 0 & 0\\
  HD 68988 & 2.1 & 0.5 & 5880 & 21 & 1.34 & 0.02 & 1.15 & 0.01 & 4.39 & 0.01 & 1.12 & 0.02 & 0 & 0\\
  HD 69830 & 10.4 & 2.5 & 5401 & 28 & 0.59 & 0.01 & 0.85 & 0.02 & 4.47 & 0.02 & 0.88 & 0.02 & 0 & 0\\
  HD 70642 & 3.6 & 0.9 & 5675 & 18 & 0.92 & 0.01 & 1.02 & 0.01 & 4.45 & 0.01 & 0.99 & 0.01 & 0 & 0\\
  HD 7199 & 9.2 & 2.5 & 5357 & 39 & 0.72 & 0.01 & 0.93 & 0.02 & 4.41 & 0.03 & 0.99 & 0.02 & 0 & 0\\
  HD 72659 & 6.4 & 0.7 & 5994 & 50 & 2.09 & 0.02 & 1.08 & 0.02 & 4.21 & 0.02 & 1.34 & 0.03 & 0 & 0\\
  HD 73256 & 4.5 & 2.3 & 5514 & 35 & 0.75 & 0.02 & 0.98 & 0.03 & 4.47 & 0.03 & 0.95 & 0.02 & 0 & 0\\
  HD 73267 & 11.8 & 1.4 & 5434 & 18 & 0.74 & 0.01 & 0.89 & 0.01 & 4.4 & 0.02 & 0.97 & 0.01 & 0 & 0\\
  HD 73526 & 7.9 & 0.5 & 5669 & 53 & 2.18 & 0.01 & 1.09 & 0.01 & 4.1 & 0.02 & 1.53 & 0.03 & 0 & 0\\
  HD 73534 & 7.1 & 0.8 & 4958 & 45 & 3.4 & 0.1 & 1.15 & 0.03 & 3.7 & 0.03 & 2.5 & 0.1 & 0 & 0\\
  HD 74156 & 4.3 & 0.6 & 6070 & 56 & 3.08 & 0.03 & 1.24 & 0.04 & 4.12 & 0.03 & 1.59 & 0.04 & 0 & 0\\
  HD 7449 & 2.2 & 1.3 & 6060 & 42 & 1.26 & 0.02 & 1.05 & 0.02 & 4.44 & 0.02 & 1.02 & 0.02 & 1 & 0\\
  HD 75289 & 1.7 & 0.4 & 6143 & 25 & 1.97 & 0.01 & 1.23 & 0.01 & 4.33 & 0.01 & 1.24 & 0.01 & 1 & 0\\
  HD 75898 & 3.8 & 0.6 & 6019 & 66 & 2.88 & 0.02 & 1.28 & 0.02 & 4.15 & 0.03 & 1.56 & 0.04 & 0 & 0\\
  HD 76700 & 6.2 & 0.9 & 5706 & 41 & 1.73 & 0.03 & 1.13 & 0.02 & 4.22 & 0.03 & 1.35 & 0.03 & 0 & 0\\
  HD 77338 & 7.8 & 3.4 & 5261 & 29 & 0.57 & 0.02 & 0.91 & 0.03 & 4.47 & 0.04 & 0.91 & 0.03 & 0 & 0\\
  HD 7924 & 5.4 & 2.6 & 5218 & 12 & 0.36 & 0.01 & 0.79 & 0.01 & 4.59 & 0.02 & 0.74 & 0.01 & 0 & 0\\
  HD 79498 & 4.2 & 0.9 & 5741 & 20 & 1.11 & 0.02 & 1.05 & 0.01 & 4.4 & 0.02 & 1.07 & 0.02 & 1 & 0\\
  HD 80606 & 2.7 & 0.7 & 5558 & 23 & 0.81 & 0.02 & 1.03 & 0.01 & 4.47 & 0.01 & 0.97 & 0.02 & 1 & 1\\
  HD 81040 & 3.6 & 1.5 & 5678 & 24 & 0.79 & 0.02 & 0.96 & 0.02 & 4.49 & 0.02 & 0.92 & 0.02 & 0 & 0\\
  HD 81688 & 6.3 & 2.9 & 4830 & 64 & 54 & 1 & 1.1 & 0.2 & 2.4 & 0.1 & 10.5 & 0.4 & 0 & 0\\
  HD 82886 & 3.3 & 0.5 & 5083 & 38 & 13.5 & 0.2 & 1.3 & 0.1 & 3.21 & 0.04 & 4.8 & 0.1 & 0 & 0\\
  HD 82943 & 3.1 & 0.4 & 5944 & 18 & 1.54 & 0.02 & 1.14 & 0.01 & 4.35 & 0.01 & 1.17 & 0.02 & 0 & 0\\
  HD 83443 & 5.2 & 1.9 & 5458 & 28 & 0.76 & 0.02 & 0.99 & 0.02 & 4.45 & 0.03 & 0.98 & 0.02 & 0 & 0\\
  HD 8535 & 3.3 & 0.5 & 6142 & 34 & 1.92 & 0.01 & 1.15 & 0.01 & 4.31 & 0.02 & 1.23 & 0.02 & 0 & 0\\
  HD 85390 & 6.8 & 2.9 & 5174 & 17 & 0.39 & 0.01 & 0.81 & 0.02 & 4.56 & 0.02 & 0.78 & 0.01 & 0 & 0\\
  HD 85512 & 8.2 & 3.0 & 4530 & 8 & 0.138 & 0.002 & 0.64 & 0.01 & 4.67 & 0.01 & 0.6 & 0.01 & 0 & 0\\
  HD 8574 & 4.4 & 0.6 & 6092 & 56 & 2.35 & 0.04 & 1.17 & 0.02 & 4.22 & 0.03 & 1.38 & 0.04 & 0 & 0\\
  HD 86081 & 5.5 & 0.9 & 5887 & 56 & 2.51 & 0.02 & 1.18 & 0.04 & 4.14 & 0.03 & 1.53 & 0.03 & 0 & 0\\
  HD 86264 & 0.8 & 0.2 & 6616 & 39 & 4.02 & 0.04 & 1.46 & 0.01 & 4.23 & 0.02 & 1.53 & 0.02 & 0 & 0\\
  HD 87883 & 7.5 & 3.8 & 4971 & 22 & 0.327 & 0.004 & 0.81 & 0.02 & 4.56 & 0.02 & 0.77 & 0.01 & 0 & 0\\
  HD 88133 & 5.7 & 0.3 & 5468 & 25 & 3.4 & 0.1 & 1.23 & 0.02 & 3.9 & 0.02 & 2.1 & 0.1 & 0 & 0\\
  HD 89307 & 4.6 & 1.7 & 6011 & 59 & 1.34 & 0.03 & 1.02 & 0.03 & 4.38 & 0.03 & 1.07 & 0.03 & 0 & 0\\
  HD 89744 & 2.5 & 0.3 & 6270 & 54 & 6.29 & 0.01 & 1.49 & 0.02 & 3.95 & 0.02 & 2.13 & 0.04 & 1 & 0\\
  HD 90156 & 5.7 & 1.7 & 5721 & 28 & 0.74 & 0.01 & 0.89 & 0.02 & 4.49 & 0.02 & 0.88 & 0.01 & 0 & 0\\
  HD 92788 & 4.2 & 1.1 & 5788 & 38 & 1.28 & 0.02 & 1.09 & 0.02 & 4.36 & 0.02 & 1.13 & 0.02 & 0 & 0\\
  HD 93083 & 6.2 & 4.4 & 5025 & 21 & 0.35 & 0.02 & 0.83 & 0.02 & 4.57 & 0.03 & 0.78 & 0.02 & 0 & 0\\
  HD 9446 & 3.7 & 2.0 & 5790 & 45 & 1.06 & 0.03 & 1.04 & 0.03 & 4.43 & 0.03 & 1.03 & 0.03 & 0 & 0\\
  HD 95089 & 3.0 & 0.2 & 4952 & 19 & 13.0 & 0.1 & 1.48 & 0.04 & 3.22 & 0.02 & 4.9 & 0.1 & 0 & 0\\
  HD 96063 & 2.9 & 0.2 & 5073 & 19 & 12.3 & 0.1 & 1.42 & 0.03 & 3.27 & 0.01 & 4.5 & 0.1 & 0 & 0\\
  HD 96127 & 7.2 & 2.1 & 3943 & 34 & 516 & 22 & 1.0 & 0.1 & 1.1 & 0.1 & 48.8 & 1.9 & 0 & 0\\
  HD 96167 & 4.7 & 0.6 & 5753 & 49 & 3.7 & 0.2 & 1.3 & 0.1 & 3.97 & 0.02 & 1.9 & 0.1 & 0 & 0\\
  HD 97658 & 9.7 & 2.8 & 5211 & 16 & 0.35 & 0.01 & 0.74 & 0.01 & 4.58 & 0.02 & 0.73 & 0.01 & 0 & 1\\
  HD 98219 & 3.2 & 0.4 & 4952 & 31 & 11.4 & 0.3 & 1.5 & 0.1 & 3.27 & 0.03 & 4.6 & 0.1 & 0 & 0\\
  HD 99109 & 6.0 & 3.0 & 5270 & 24 & 0.56 & 0.02 & 0.93 & 0.02 & 4.49 & 0.03 & 0.9 & 0.03 & 0 & 0\\
  HD 99492 & 8.1 & 3.2 & 4917 & 21 & 0.33 & 0.01 & 0.82 & 0.02 & 4.55 & 0.02 & 0.79 & 0.02 & 1 & 0\\
  HD 99706 & 2.9 & 0.2 & 4847 & 17 & 15.6 & 0.1 & 1.53 & 0.03 & 3.12 & 0.02 & 5.6 & 0.1 & 0 & 0\\
  HIP 14810 & 7.1 & 2.1 & 5570 & 47 & 0.99 & 0.03 & 1.00 & 0.02 & 4.37 & 0.04 & 1.07 & 0.04 & 0 & 0\\
  HIP 5158 & 4.5 & 3.2 & 4571 & 14 & 0.19 & 0.01 & 0.75 & 0.01 & 4.63 & 0.02 & 0.69 & 0.02 & 0 & 0\\
  HIP 57050 & 8.8 & 3.6 & 3542 & 2 & 0.0068 & 2.0E-4 & 0.2 & 0.0 & 5.05 & 0.01 & 0.219 & 0.003 & 0 & 0\\
  HIP 57274 & 8.9 & 2.1 & 4636 & 35 & 0.2 & 0.01 & 0.72 & 0.01 & 4.61 & 0.01 & 0.69 & 0.03 & 0 & 0\\
  kappa CrB & 3.2 & 0.4 & 4899 & 30 & 11.8 & 0.2 & 1.5 & 0.1 & 3.24 & 0.03 & 4.8 & 0.1 & 0 & 0\\
  KELT-3 & 1.8 & 1.2 & 6413 & 55 & 3.3 & 0.1 & 1.3 & 0.02 & 4.2 & 0.01 & 1.5 & 0.1 & 0 & 1\\
  KELT-6 & 2.6 & 2.8 & 6267 & 46 & 3.8 & 0.6 & 1.2 & 0.1 & 4.1 & 0.1 & 1.7 & 0.1 & 0 & 1\\
  Kepler-21 & 3.7 & 0.4 & 6264 & 60 & 5.20 & 0.04 & 1.31 & 0.03 & 3.97 & 0.02 & 1.94 & 0.04 & 0 & 1\\
  Kepler-37 & 3.3 & 0.6 & 5630 & 9 & 0.559 & 0.002 & 0.85 & 0.01 & 4.57 & 0.01 & 0.787 & 0.004 & 0 & 1\\
  Kepler-68 & 6.1 & 0.5 & 5868 & 32 & 1.62 & 0.03 & 1.07 & 0.01 & 4.28 & 0.01 & 1.23 & 0.03 & 0 & 1\\
  mu Ara & 5.4 & 0.7 & 5817 & 44 & 1.78 & 0.04 & 1.14 & 0.02 & 4.25 & 0.03 & 1.32 & 0.04 & 0 & 0\\
  NGC 2423 3 & 6.7 & 3.0 & 4446 & 26 & 73 & 14 & 1.2 & 0.2 & 2.19 & 0.01 & 14.4 & 1.6 & 0 & 0\\
  Qatar-1 & 6.9 & 3.8 & 4730 & 15 & 0.24 & 0.01 & 0.78 & 0.02 & 4.59 & 0.02 & 0.73 & 0.01 & 0 & 1\\
  tau Boo & 1.8 & 0.4 & 6408 & 45 & 3.11 & 0.04 & 1.33 & 0.01 & 4.24 & 0.02 & 1.43 & 0.03 & 1 & 0\\
  TrES-1 & 9.5 & 3.1 & 5492 & 57 & 0.77 & 0.02 & 0.91 & 0.03 & 4.42 & 0.04 & 0.97 & 0.03 & 0 & 1\\
  TrES-2 & 5.0 & 1.0 & 5958 & 65 & 1.7 & 0.1 & 1.09 & 0.03 & 4.3 & 0.01 & 1.2 & 0.1 & 0 & 1\\
  TrES-3 & 7.9 & 1.4 & 5614 & 29 & 0.77 & 0.01 & 0.9 & 0.02 & 4.45 & 0.01 & 0.93 & 0.02 & 0 & 1\\
  TrES-4 & 2.1 & 0.1 & 6327 & 25 & 3.42 & 0.04 & 1.37 & 0.01 & 4.192 & 0.004 & 1.54 & 0.02 & 0 & 1\\
  TrES-5 & 7.5 & 3.7 & 5087 & 31 & 0.41 & 0.01 & 0.85 & 0.03 & 4.53 & 0.03 & 0.82 & 0.02 & 0 & 1\\
  WASP-11 & 5.7 & 2.1 & 4917 & 18 & 0.29 & 0.01 & 0.79 & 0.01 & 4.59 & 0.01 & 0.74 & 0.02 & 0 & 1\\
  WASP-14 & 1.5 & 1.4 & 6454 & 52 & 4.2 & 0.5 & 1.33 & 0.04 & 4.13 & 0.03 & 1.7 & 0.1 & 0 & 1\\
  WASP-16 & 8.5 & 1.2 & 5633 & 49 & 1.09 & 0.04 & 0.98 & 0.02 & 4.34 & 0.01 & 1.1 & 0.04 & 0 & 1\\
  WASP-18 & 0.9 & 0.2 & 6167 & 7 & 1.7 & 0.04 & 1.2 & 0.001 & 4.39 & 0.01 & 1.15 & 0.02 & 0 & 1\\
  WASP-19 & 3.6 & 1.8 & 5526 & 39 & 0.76 & 0.01 & 1.00 & 0.02 & 4.47 & 0.02 & 0.95 & 0.02 & 0 & 1\\
  WASP-2 & 0.7 & 0.3 & 5345 & 8 & 0.47 & 0.01 & 0.9 & 0.001 & 4.581 & 0.004 & 0.8 & 0.01 & 0 & 1\\
  WASP-21 & 3.7 & 0.7 & 6123 & 38 & 1.43 & 0.03 & 1.03 & 0.02 & 4.39 & 0.01 & 1.06 & 0.02 & 0 & 1\\
  WASP-25 & 3.2 & 1.2 & 5582 & 21 & 0.72 & 0.02 & 0.97 & 0.01 & 4.5 & 0.01 & 0.91 & 0.02 & 0 & 1\\
  WASP-26 & 3.1 & 0.4 & 5881 & 13 & 1.26 & 0.01 & 1.09 & 0.01 & 4.4 & 0.01 & 1.08 & 0.01 & 0 & 1\\
  WASP-34 & 6.8 & 1.3 & 5758 & 55 & 1.19 & 0.03 & 1.01 & 0.02 & 4.35 & 0.02 & 1.1 & 0.04 & 0 & 1\\
  WASP-37 & 7.7 & 1.4 & 5956 & 55 & 1.1 & 0.04 & 0.9 & 0.02 & 4.39 & 0.02 & 0.99 & 0.04 & 0 & 1\\
  WASP-39 & 6.7 & 2.4 & 5466 & 34 & 0.57 & 0.01 & 0.86 & 0.02 & 4.51 & 0.02 & 0.85 & 0.01 & 0 & 1\\
  WASP-4 & 5.5 & 2.0 & 5435 & 31 & 0.6 & 0.01 & 0.91 & 0.02 & 4.51 & 0.02 & 0.88 & 0.01 & 0 & 1\\
  WASP-41 & 3.9 & 1.0 & 5555 & 27 & 0.7 & 0.02 & 0.96 & 0.01 & 4.5 & 0.01 & 0.91 & 0.02 & 0 & 1\\
  WASP-43 & 2.4 & 2.3 & 4756 & 12 & 0.215 & 0.002 & 0.75 & 0.01 & 4.64 & 0.01 & 0.68 & 0.01 & 0 & 1\\
  WASP-44 & 6.8 & 2.8 & 5402 & 31 & 0.6 & 0.02 & 0.9 & 0.02 & 4.49 & 0.03 & 0.89 & 0.02 & 0 & 1\\
  WASP-48 & 9.1 & 0.8 & 5770 & 53 & 2.6 & 0.2 & 1.01 & 0.02 & 4.03 & 0.01 & 1.6 & 0.1 & 0 & 1\\
  WASP-5 & 2.5 & 0.7 & 5611 & 14 & 0.79 & 0.02 & 1.01 & 0.01 & 4.49 & 0.01 & 0.94 & 0.01 & 0 & 1\\
  WASP-52 & 3.8 & 2.6 & 5077 & 14 & 0.348 & 0.004 & 0.83 & 0.02 & 4.58 & 0.02 & 0.76 & 0.01 & 0 & 1\\
  WASP-54 & 3.3 & 0.4 & 6090 & 26 & 1.61 & 0.02 & 1.1 & 0.01 & 4.36 & 0.01 & 1.14 & 0.02 & 0 & 1\\
  WASP-58 & 12.6 & 0.1 & 5874 & 5 & 1.23 & 0.01 & 0.839 & 0.002 & 4.2936 & 1.0E-4 & 1.073 & 0.005 & 0 & 1\\
  WASP-60 & 8.8 & 1.4 & 5730 & 52 & 1.09 & 0.03 & 0.95 & 0.02 & 4.36 & 0.02 & 1.06 & 0.04 & 0 & 1\\
  WASP-75 & 2.9 & 0.2 & 5862 & 9 & 1.1 & 0.01 & 1.051 & 0.004 & 4.435 & 0.002 & 1.02 & 0.01 & 0 & 1\\
  WASP-8 & 4.2 & 1.1 & 5632 & 34 & 0.96 & 0.02 & 1.04 & 0.02 & 4.42 & 0.01 & 1.03 & 0.02 & 1 & 1\\
  XO-2 & 9.6 & 0.9 & 5474 & 54 & 1.7 & 0.1 & 1.04 & 0.03 & 4.13 & 0.01 & 1.4 & 0.1 & 1 & 1\\
  XO-3 & 1.7 & 1.0 & 6685 & 45 & 3.7 & 0.1 & 1.26 & 0.01 & 4.22 & 0.01 & 1.43 & 0.04 & 0 & 1\\
  XO-5 & 12.4 & 0.6 & 5452 & 12 & 0.794 & 0.004 & 0.89 & 0.01 & 4.38 & 0.01 & 1.00 & 0.01 & 0 & 1\\
\end{longtable}
\end{longtab}
%
}

\bibliography{biblio}
\bibliographystyle{aa}

\end{document}